\documentclass[useAMS,usenatbib]{mn2e}
\usepackage{txfonts}
\usepackage{graphicx}
\usepackage{natbib}

\def\del#1{{}}

\newcommand{\ltsima}{$\; \buildrel < \over \sim \;$}
\newcommand{\lsim}{\lower.5ex\hbox{\ltsima}}
\newcommand{\gtsima}{$\; \buildrel > \over \sim \;$}
\newcommand{\gsim}{\lower.5ex\hbox{\gtsima}}

\newcommand{\bra}{\langle}
\newcommand{\ket}{\rangle}

\newcommand{\planck}{{\em Planck}}
\newcommand{\plancks}{{\em Planck}'s }

\newcommand{\mK}{{\em m}K}

\newcommand{\nK}{{\em n}K}
\newcommand{\hfi}{{\em HFI}}
\newcommand{\lfi}{{\em LFI}}

\newcommand{\dd}{\mathrm{d}}
\newcommand{\e}{\mathrm{e}}

\title[Detecting SZ-Clusters with \planck]
{Detecting Sunyaev-Zel'dovich clusters with \planck:\\ II. Foreground components and optimised filtering schemes}
\author[B. M. Sch\"afer, C. Pfrommer, R. M. Hell and M. Bartelmann]
{B. M. Sch\"afer$^{1}$\thanks{e-mail: spirou@mpa-garching.mpg.de (BMS);
pfrommer@mpa-garching.mpg.de (CP); reinhard@mpa-garching.mpg.de (RMH); mbartelmann@ita.uni-heidelberg.de (MB)},
C. Pfrommer$^{1}$\footnotemark[1], R. M. Hell$^{1}$\footnotemark[1] and M. Bartelmann$^{2}$\footnotemark[1]\\
$^1$Max-Planck-Institut f\"ur Astrophysik, Karl-Schwarzschild-Stra{\ss}e 1, Postfach 1317, 85741 Garching, Germany\\
$^2$Institut f\"ur theoretische Astrophysik, Tiergartenstra{\ss}e 15, 69121 Heidelberg, Germany}

\begin{document}
\pagerange{\pageref{firstpage}--\pageref{lastpage}}
\pubyear{2003}
\maketitle
\label{firstpage}

\begin{abstract}
The \planck~mission is the most sensitive all-sky CMB experiment currently planned. The High Frequency Instrument (\hfi) 
will be especially suited to observe clusters of galaxies by their thermal Sunyaev-Zel'dovich (SZ) effect. In order to 
assess \plancks~SZ-capabilities in the presence of spurious signals, a simulation is presented that combines maps 
of the thermal and kinetic SZ-effects with a realisation of the cosmic microwave background (CMB), in addition to 
Galactic foregrounds (synchrotron emission, free-free emission, thermal emission from dust, CO-line radiation) as well 
as the sub-millimetric emission from celestial bodies of our Solar system. Additionally, observational issues such as 
the finite angular resolution and spatially non-uniform instrumental noise of \plancks sky maps are taken into account, 
yielding a set of all-sky flux maps, the auto-correlation and cross-correlation properties of which are examined 
in detail. In the second part of the paper, filtering schemes based on scale-adaptive and matched filtering are extended to 
spherical data sets, that enable the amplification of the weak SZ-signal in the presence of all contaminations stated above. 
The theory of scale-adaptive and matched filtering in the framework of spherical maps is developed, the resulting filter kernel 
shapes are discussed and their functionality is verified.
\end{abstract}

\begin{keywords}
galaxies: clusters: general, cosmology: cosmic microwave background, methods: numerical, space vehicles: {\em 
Planck}
\end{keywords}

\section{Introduction}\label{sect_intro}
The Sunyaev-Zel'dovich (SZ) effect \citep{1972SZorig, 1980ARA&A..18..537S, 1995ARA&A..33..541R, 1993birkinshaw} is the 
most important extragalactic source of secondary anisotropies in the CMB sky. The thermal SZ-effect is explained by the fact 
that CMB photons are put in thermal contact with electrons of the hot intra-cluster medium (ICM) by Compton-interactions which 
causes a transfer of energy from the ICM to the CMB. Because of the smallness of the Thompson cross-section and of the 
diluteness of the ICM this transfer of thermal energy is small. In the direction of a cluster, low-energetic photons with 
frequencies below $\nu=217$~GHz are removed from the line-of-sight. At frequencies above $\nu=217$~GHz CMB photons are 
scattered into the line-of-sight, causing a distinct modulation of the CMB surface brightness as a function of observing 
frequency, which enables the detection of clusters of galaxies in microwave data.

In contrast, in the kinetic effect it is the peculiar motion of a cluster along the line of sight relative to the CMB 
frame that induces CMB surface brightness fluctuations. The peculiar motion of the cluster causes the CMB to be anisotropic 
in the cluster frame. Due to this symmetry breaking of the scattering geometry, photons scattered into the line-of-sight 
are shifted in frequency, namely to higher frequencies, if the cluster is moving towards the observer.

The \planck-mission will be especially suited to detect SZ-clusters due to its sensitivity, its spectroscopic capabilities, 
sky coverage and spatial resolution. It is expected to yield a cluster catalogue containing $\simeq10^4$ entries. Extensive 
literature exists on the topic, but so far the influence of foregrounds and details of \plancks instrumentation and data 
aquisition have not been thoroughly addressed. In this work we aim at modelling the astrophysical and instrumental issues 
connected to the observation of SZ-clusters as exhaustively as possible: A simulation is presented that combines realistic maps 
of both SZ-effects with a realisation of the CMB, with four different Galactic foreground components (thermal dust, free-free 
emission, synchrotron emission and emission from rotational transitions of CO molecules), with maps containing the 
sub-millimetric emission from planets and asteroids of the Solar system and with instrumental noise. \plancks frequency 
response and beam shapes are modelled conforming to the present knowledge of \plancks receivers and its optical system. In 
order to extract the SZ-cluster signal, filtering schemes based on matched and scale-adaptive filtering are extended to 
spherical data sets. 

In contrast to the recent work by \citet{2004astro.ph..6190G}, our SZ-simulation does not rely on idealised scaling relations 
and futhermore, takes account of the cluster's morphological variety. The Galactic foregrounds are modelled in concordance with 
WMAP observations \citep[see][]{2003ApJS..148...97B}, which constitues an improvement over the simplifying assumptions made by 
Geisb{\"u}sch et al. In addition, instrumentation issues such as non-isotropic detector noise are properly incorporated 
into the simulation. The filter scheme employed in the paper by Geisb{\"u}sch et al. is the harmonic-space maximum entropy 
method introduced by \citet{2002MNRAS.336...97S} which assumes approximate prior knowledge of the emission component's power 
spectra. Its computational demand is much higher than matched and scale-adaptive filtering: In fact, the computations presented 
in this work can be run on a notebook-class computer. 

The paper is structured as follows: After a brief recapitulation of the SZ-effect in Sect.~\ref{sect_szdef}, the 
\planck-satellite and instrumental issues connected to observation of CMB anisotropies are decribed in Sect.~\ref{sect_planck}. 
The foreground emission components are introduced in Sect.~\ref{sect_foreground}. The steps in the simulation of flux maps 
for the various \planck-channels are described and their correlation properties are examined in Sect.~\ref{sect_plancksim}. The 
theory of matched and scale-adaptive filtering is extended to spherical data sets and the resulting filter kernel shapes are in 
detail discussed in Sect.~\ref{sect_filtering}. A summary in Sect.~\ref{sect_summary} concludes the paper.

Throughout the paper, the cosmological model assumed is the standard \mbox{$\Lambda$CDM} cosmology, which has recently 
been supported by observations of the WMAP satellite \citep{2003astro.ph..2209S}. Parameter values have been chosen as 
$\Omega_\mathrm{M} = 0.3$, $\Omega_\Lambda =0.7$, $H_0 = 100\,h\,\mbox{km~}\mbox{s}^{-1}\mbox{ Mpc}^{-1}$ with $h = 0.7$, 
$\Omega_\mathrm{B} = 0.04$, $n_\mathrm{s} =1$ and $\sigma_8=0.9$.

\section{Sunyaev-Zel'dovich definitions}\label{sect_szdef}
The Sunyaev-Zel'dovich effects are the most important extragalactic sources of secondary anisotropies in the CMB. Inverse 
Compton scattering of CMB photons with electrons of the ionised ICM gives rise to these effects and induce surface brightness
fluctuations of the CMB sky, either because of the thermal motion of the ICM electrons (thermal SZ-effect) or because of 
the bulk motion of the cluster itself relative to the comoving CMB-frame along the line-of-sight (kinetic SZ-effect).

The relative change $\Delta T/T$ in thermodynamic CMB temperature at position $\bmath{\theta}$ as a function of
dimensionless frequency $x=h\nu /(k_B T_\mathrm{CMB})$ due to the thermal SZ-effect is given by:
\begin{eqnarray}
\frac{\Delta T}{T}(\bmath{\theta}) & = & y(\bmath{\theta})\,\left(x\frac{e^x+1}{e^x-1}-4\right)\mbox{ with }\\
y(\bmath{\theta}) & = & \frac{\sigma_\mathrm{T} k_B}{m_\e c^2}\int\dd 
l\:n_\e(\bmath{\theta},l)T_\e(\bmath{\theta},l)\mbox{,}
\label{sz_temp_decr}
\end{eqnarray}
where the amplitude $y$ of the thermal SZ-effect is commonly known as the thermal Comptonisation parameter, that
itself is defined as the line-of-sight integral of the temperature weighted thermal electron density. $m_\e$, $c$,
$k_B$ and $\sigma_\mathrm{T}$ denote electron mass, speed of light, Boltzmann's constant and the Thompson cross section,
respectively. The kinetic SZ-effect arises due to the motion of the cluster parallel to the line of sight relative to 
the CMB-frame:
\begin{equation}
\frac{\Delta T}{T}(\bmath{\theta}) = -w(\bmath{\theta})\mbox{ with }
w(\bmath{\theta}) = \frac{\sigma_\mathrm{T}}{c}\int\dd l\:n_\e(\bmath{\theta,l})\upsilon_r(\bmath{\theta},l)\mbox{.}
\end{equation}
Here, $\upsilon_r$ is the radial component of the cluster's velocity. The convention is such that $\upsilon_r<0$, if the 
cluster is moving towards the observer. In this case, the CMB temperature is increased. In analogy, the quantity $w$ is refered 
toas the kinetic Comptonisation. The SZ-observables are the line-of-sight Comptonisations integrated over the solid angle 
subtended by the cluster. The quantities $\mathcal{Y}$ and $\mathcal{W}$ are refered to as the integrated thermal and 
kinetic Comptonisations, respectively:
\begin{eqnarray}
\mathcal{Y} & = & \int\dd\Omega\: y(\bmath{\theta}) = 
d_A^{-2}(z)\cdot\frac{\sigma_\mathrm{T} k_B}{m_e c^2}\:\int\dd V\:n_e T_e\\
\mathcal{W} & = & \int\dd\Omega\: w(\bmath{\theta}) = 
d_A^{-2}(z)\cdot\frac{\sigma_\mathrm{T}}{c}\:\int\dd V\:n_e \upsilon_r
\end{eqnarray}
Here, $d_A(z)$ denotes the angular diameter distance of a cluster situated at redshift $z$.

\section{Submillimetric observations with \planck}\label{sect_planck}
The \planck-mission\footnote{\tt http://planck.mpa-garching.mpg.de/}$^{,}$\footnote{\tt http://astro.estec.esa.nl/Planck/} will 
perform a polarisation sensitive survey of the complete microwave sky in nine observing frequencies from the Lagrange point 
$L_2$ in the Sun-Earth system. It will observe at angular resolutions of up to $5\farcm0$ in the best channels and will 
achieve micro-Kelvin sensitivity relying on bolometric receivers \citep[high frequency instrument \hfi, described 
in][]{lit_hfi} and on high electron mobility transistors \citep[low frequency instrument \lfi, 
see][]{2003astro.ph..4137V,2000ApL&C..37..171B}. The main characteristics are summarised in Table~\ref{table_planck_channel}.
\plancks beam characteristics are given Sect.~\ref{planck_beamshape} and the scanning strategy and the simulation of spatially 
non-uniform detector noise is outlined in Sect.~\ref{planck_scannoise}.

\begin{table*}\vspace{-0.1cm}
\begin{center}
\begin{tabular}{lrrrrrrrrr}
\hline\hline
\vphantom{\Large A}%
\planck~ channel	& 1 & 2 & 3 & 4 & 5 & 6 & 7 & 8 & 9 \\
\hline
\vphantom{\Large A}%
centre frequency $\nu_0$				
& 30~GHz   & 44~GHz   & 70~GHz   & 100~GHz & 143~GHz & 217~GHz & 353~GHz & 545~GHz & 857~GHz 		\\
frequency window $\Delta\nu$			
& 3.0~GHz & 4.4~GHz & 7.0~GHz & 16.7~GHz & 23.8~GHz & 36.2~GHz & 58.8~GHz & 90.7~GHz & 142.8~GHz	\\
resolution $\Delta\theta$ (FWHM) & $33\farcm4$ & $26\farcm8$ & $13\farcm1$ & $9\farcm2$ &$7\farcm1$ & $5\farcm0$ & 
$5\farcm0$ & $5\farcm0$ & $5\farcm0$\\
noise level $\sigma_\mathrm{N}$	& 1.01~\mK & 0.49~\mK & 0.29~\mK & 5.67~\mK & 
4.89~\mK & 6.05~\mK & 6.80~\mK & 3.08~\mK & 4.49~\mK \\
\hline
thermal SZ-flux $\bra S_\mathcal{Y}\ket$	
& -12.2~Jy & -24.8~Jy & -53.6~Jy & -82.1~Jy & -88.8~Jy &  -0.7~Jy & 146.0~Jy & 76.8~Jy  & 5.4~Jy	\\
kinetic SZ-flux $\bra S_\mathcal{W}\ket$
& 6.2~Jy   & 13.1~Jy  & 30.6~Jy  &  55.0~Jy &  86.9~Jy & 110.0~Jy &  69.1~Jy & 15.0~Jy  & 0.5~Jy	\\
antenna temperature $\Delta T_\mathcal{Y}$ 	
& -440~\nK & -417~\nK & -356~\nK & -267~\nK & -141~\nK &  -0.5~\nK&   38~\nK &  8.4~\nK & 0.2~\nK	\\
antenna temperature $\Delta T_\mathcal{W}$ 	
& 226~\nK  & 220~\nK  & 204~\nK  &  179~\nK &  138~\nK &    76~\nK&   18~\nK &  1.6~\nK & 0.02~\nK	\\
\hline
\end{tabular}
\end{center}
\caption{Characteristics of \plancks \lfi-receivers (column 1-3) and \hfi-bolometers (column 4-9): centre frequency $\nu_0$, 
frequency window $\Delta\nu$ as defined in eqns.~(\ref{eqn_tlm_exp}) and (\ref{eq_freq_resp}), angular resolution 
$\Delta\theta$ stated in FWHM, effective noise level $\sigma_\mathrm{N}$, fluxes $\bra S_\mathcal{Y}\ket$ and $\bra 
S_\mathcal{W}\ket$ generated by the respective Comptonisation of $\mathcal{Y}=\mathcal{W}=1~\mathrm{arcmin}^2$ and the 
corresponding changes in antenna temperature $\Delta T_\mathcal{Y}$ and $\Delta T_\mathcal{W}$. Due to \plancks symmetric 
frequency response window, the thermal SZ-effect does not vanish entirely at $\nu=217$~GHz.}
\label{table_planck_channel}
\end{table*}

\subsection{Beam shapes}\label{planck_beamshape}
\label{sect_beam}
The beam shapes of \planck~are well described by azimuthally symmetric Gaussians $b(\theta) = 
\frac{1}{2\pi\sigma_\theta^2}\exp\left(-\frac{\theta^2}{2\sigma_\theta^2}\right)$ with $\sigma_\theta = 
\frac{\Delta\theta}{\sqrt{8\ln(2)}}$. The residuals from the ideal Gaussian shape (ellipticity, higher order distortions, 
diffraction rings, far-side lobes, pick-up of stray-light) are expected not to exceed the percent level and are neglected for 
the purpose of this work. Table~\ref{table_planck_channel} gives the angular resolution $\Delta\theta$ in terms of FWHM of each 
\planck-channel for reference.

\subsection{Scanning strategy and noise-equivalent maps}\label{planck_scannoise}
CMB observations by \planck~will proceed in great circles fixed on the ecliptic poles. A single scan will start at 
the North ecliptic pole, will follow a meridian to the South ecliptic pole and back to the North ecliptic pole by 
following the antipodal meridian. Such a scan will last one minute and will be repeated sixty times. After that, the 
rotation axis will be shifted in a precessional motion for $2\farcm5$ (approximately half a beam diameter) and the scan 
repeated. In this way, the entire sky is mapped once in 180 days.

Fourier transform of the noise time series of \plancks receivers yields a noise power spectrum $P(f)$ of the shape 
\begin{equation}
P(f) = \sigma_\mathrm{N}^2\cdot\left[1 + \left(\frac{f}{f_\mathrm{knee}}\right)^{-\alpha}\right]\mbox{,}
\end{equation}
i.e. the noise consists of two components: a power law component in frequency $f$, decribed by the spectral index $\alpha$ that 
assuming values $0\leq\alpha\leq 2$ and a white noise component, smoothly joined at the frequency $f_\mathrm{knee}$. 

The $f^{-\alpha}$-part of the noise spectrum originates from zero point drifts of the detector gain on large time scales.
This power law component exhibits low-frequency variations that lead to the typical stripe pattern in simulated 
\planck-maps due to the scanning strategy \citep{1999A&AS..140..383M}. Algorithms for destriping the maps are a current 
research topic (for example, the {\tt Mirage}-algorithm proposed by \citet{2004astro.ph..1505Y},
{\tt MAPCUMBA} by \citet{2001A&A...374..358D} and the max-likelihood algorithm by \citet{2001A&A...372..346N}), but it can be 
expected that the destriping can be done very efficiently such that the remaining noise largely consists of uncorrelated pixel 
noise.

In order to incorporate uncorrelated pixel noise into the simulation, a set of maps has been construced, where at each pixel a 
number from a Gaussian distribution with width $\sigma_\mathrm{N}$ has been drawn. For \plancks \hfi-receivers, the 
rms-fluctuations $\sigma_\mathrm{N}$ in antenna temperature can be calculated from the noise equivalent power NEP and the 
sampling frequency $\nu_\mathrm{sampling}=200$~Hz via:
\begin{equation}
\sigma_\mathrm{N} = \frac{2~\mathrm{NEP}\sqrt{\nu_\mathrm{sampling}}}{k_B \Delta\nu}
\quad\mbox{(\hfi)}
\label{eqn_hfi_noise}
\end{equation}

Alternatively, for \plancks \lfi-receivers, the rms-fluctuations $\sigma_\mathrm{N}$ in antenna temperature are given by:
\begin{equation}
\sigma_\mathrm{N} = \sqrt{2}\frac{T_\mathrm{noise} + T_\mathrm{CMB}}{\sqrt{\Delta\nu/\nu_\mathrm{sampling}}}
\quad\mbox{(\lfi)}
\label{eqn_lfi_noise}
\end{equation}
Values for $T_\mathrm{noise}$ and NEP can be obtained from \plancks simulation pipeline manual. The resulting effective noise 
level for all \planck~channels for a single observation of a pixel is given in Table~\ref{table_planck_channel}. The formulae 
and respective parameters are taken from the-planck simulation manual, available via \plancks {\tt LiveLink}.

The rms-fluctuations $\sigma_\mathrm{N}$ in antenna temperature have to be scaled by $\sqrt{n_\mathrm{det}}$  (assuming 
Poissonian statistics), where $n_\mathrm{det}$ denotes the number of redundant receivers per channel, because they 
provide independent surveys of the microwave sky. 

\begin{figure}
\begin{center}
\resizebox{5cm}{!}{\includegraphics{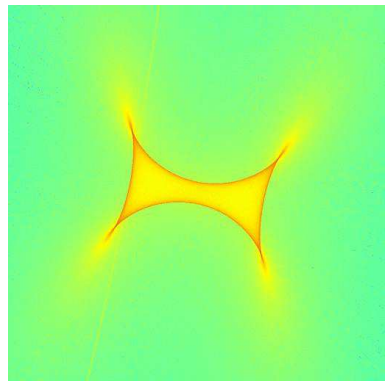}}
\end{center}
\caption{Exposure map ($\mathrm{side~length}\simeq70\degr$) of a single $\nu=353$~GHz-receiver at the North ecliptic pole in 
logarithmic shading: The displacement of the receiver with respect to the optical axis causes the observational rings not 
to overlap exactly at the pole, but gives rise to the lozenge-shaped pattern in the sky-coverage map. On average, the pixels 
inside the lozenge are observed roughly 100 times, compared to $\sim 20$ times outside. Pixels on the edges of the lozenge are 
observed a few thousand times. The best observed pixels are situated on the tips of the lozenge, where values as high as 
$2\cdot10^4$ are attained. The numbers correspond to the planned mission lifetime of 1 year. The faint diagonal tangential line 
on the left side is caused by 2008's being a leap year: The mapping of the entire sky would be completed in 365 days, but there 
is an additional day available.}
\label{figure_exposure_map}
\end{figure}

From simulated scanning paths it is possible to derive an exposure map using the {\tt simmission}- and {\tt 
multimod}-utilities. An example of such an exposure map in the vicinity of the North ecliptic pole is given in 
Fig.~\ref{figure_exposure_map}. Using the number of observations $n_\mathrm{obs}$ per pixel, it is possible to scale down the 
noise amplitudes by $\sqrt{n_\mathrm{obs}}$ and to obtain a realistic noise map for each channel. Here, we apply the 
simplification that all detectors of a given channel are arranged collinearly. In this case, the exposure maps will have sharp 
transitions from well-observed regions around the ecliptic poles to the region around the ecliptic equator. In real 
observations these transitions will be smoothed out due slight displacements of the optical axes among each other which causes 
the effective exposure pattern to be a superposition of rotated and distorted single-receiver exposure patterns.

\section{Foreground emission components}\label{sect_foreground}
The observation of the CMB and of SZ-clusters is seriously impeded by various Galactic foregrounds and by the thermal 
emission of celestial bodies of our Solar system. In order to describe these emission components, template maps from 
microwave surveys are used. \citet{1999NewA....4..443B} give a comprehensive review for the foreground components 
relevant for the \planck~mission. As foreground components we include thermal emission from dust in the Galactic 
plane (Sect.~\ref{foreground_dust}), Galactic synchrotron (Sect.~\ref{foreground_synchro}) and free-free emission 
(Sect.~\ref{foreground_freefree}), line emission from rotational transitions of carbon monoxide molecules in giant 
molecular clouds (Sect.~\ref{foreground_co}), sub-millimetric emission from planets (Sect.~\ref{foreground_planets}) 
and from minor bodies of the Solar system (Sect.~\ref{foreground_asteroids}). Foreground components omitted at this stage are 
discussed in Sect.~\ref{foreground_omitted}.

In this work, no attempt is made at modelling the interactions between various foreground components because of poorly known 
parameters such as the spatial arrangement along the line-of-sight of the emitting and absorbing components. Exemplarily, the 
reader is referred to \citet{2003ApJS..146..407F}, where the absorption of Galactic free-free emission by dust is discussed.

\subsection{Galactic dust emission}\label{foreground_dust}
At frequencies above $\sim$~100~GHz, the thermal emission from dust in the disk of the Milky Way is the most prominent 
feature in the microwave sky. Considerable effort has been undertaken to model the thermal emission 
from Galactic dust \citep{1997AAS...191.8704S,1998ApJ...500..525S,1999ApJ...524..867F,2000ApJ...544...81F}. The thermal 
dust emission is restricted to low Galactic latitudes and the thin disk is easily discernible.

\begin{figure}
\resizebox{\hsize}{!}{\includegraphics{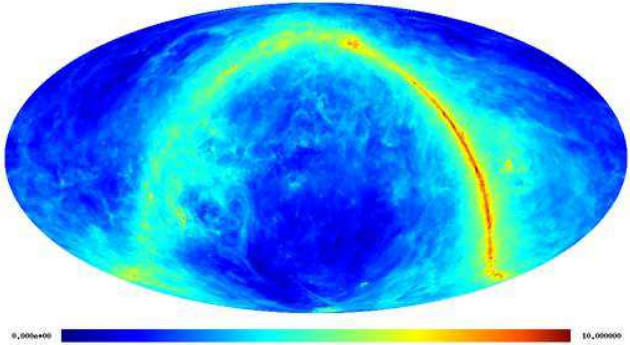}}
\caption{Sky map of the antenna temperature increase caused by dust emission in the $\nu=100$~GHz-channel: The shading is 
proportional to $\mathrm{arsinh}(T_A(\nu=100~\mathrm{GHz}) / \mathrm{\umu K})$. Ecliptic coordinates have been chosen. This map 
has been derived from the dust-template map provided by \citet{1998ApJ...500..525S}.}
\label{figure_dustmap}
\end{figure}

The input template map (see Fig.~\ref{figure_dustmap}) is derived from an observation at a wavelength of 
$\lambda=100~\umu\mathrm{m}$, i.e. $\nu_0=3~\mathrm{THz}$. Its amplitudes $A_\mathrm{dust}$ are given in MJy/sr, 
which are extrapolated to the actual frequency channels of \planck~using a two-component model suggested by C. Baccigalupi 
(personal communication). Despite the fact that the dust is expected to spread over a large range of temperatures, the model 
reproduces the thermal emission remarkably well. This model yields for the flux $S_\mathrm{dust}(\nu)$:

\begin{equation}
S_\mathrm{dust}(\nu) = 
\frac{f_1 q\cdot\left(\frac{\nu}{\nu_0}\right)^{\alpha_1} B(\nu,T_1) + 
f_2\cdot\left(\frac{\nu}{\nu_0}\right)^{\alpha_2} B(\nu,T_2)}{f_1 q B(\nu_0,T_1) + f_2 B(\nu_0,T_2)}\cdot 
A_\mathrm{dust}\mbox{.}
\end{equation}
The choice of parameters used is: $f_1 = 0.0363$, $f_2 = 1-f_1$, $\alpha_1=1.67$, $\alpha_2=2.70$, $q=13.0$. The two 
dust temperatures are $T_1 = 9.4$~K and $T_2 = 16.2$~K. The function $B(\nu,T)$ denotes the Planckian emission-law:
\begin{equation}
B(\nu,T) = \frac{2h}{c^2}\cdot\frac{\nu^3}{\exp(h\nu/k_B T)-1}\mbox{.}
\end{equation}

\subsection{Galactic synchrotron emission}\label{foreground_synchro}
\begin{figure}
\resizebox{\hsize}{!}{\includegraphics{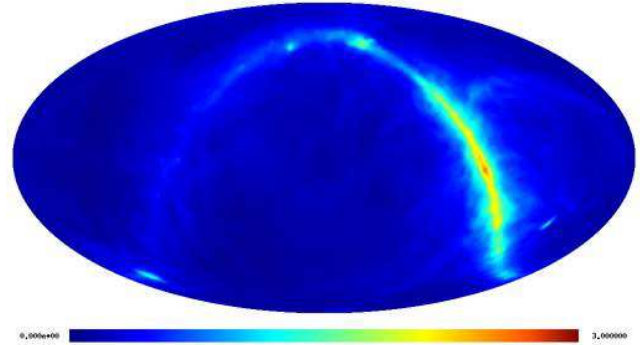}}
\caption{Sky map of the antenna temperature increase caused by synchrotron emission in the $\nu=100$~GHz-channel in ecliptic 
coordinates: The shading is proportional to $\mathrm{arsinh}(T_A(\nu=100~\mathrm{GHz}) / \mathrm{\umu K})$. The survey 
undertaken by \citet{1982A&AS...47....1H} was used to construct this template.}
\label{figure_synchromap}
\end{figure}

Relativistic electrons of the interstellar medium produce synchrotron radiation by spiralling around magnetic field 
lines, which impedes CMB observations most strongly at frequencies below 100~GHz. The synchrotron emission reaches out 
to high Galactic latitude and is an important ingredient for modelling foreground emission in microwave observations. An 
all-sky survey at an observing frequency of 408~MHz has been compiled by \citet{1981A&A...100..209H, 1982A&AS...47....1H} and 
adopted for usage with \planck~by \citet{2002A&A...387...82G} (see Fig.~\ref{figure_synchromap}). The average angular 
resolution of this survey is $0\fdg85$ (FWHM).

Recent observations with WMAP \citep{2003ApJS..148...97B} indicate that the spectral slope of the synchrotron emission 
changes dramatically from $\gamma = -0.75$ at frequencies below 22~GHz to $\gamma = -1.25$ above 22~GHz. Theoretically, this 
may be explained by a momentum-dependent diffusion coefficient for cosmic ray electrons. In order to take account of this 
spectral steepening, the amplitudes $A_\mathrm{synchro}$ are multiplied with a prefactor in order to obtain the synchrotron 
fluxes at $\nu=22~\mathrm{GHz}$. This value is then extrapolated to \plancks observing frequencies with a spectral index of 
$\gamma = -1.25$: The amplitudes $A_\mathrm{synchro}$ of the input map are given in units of MJy/sr, and for the flux 
$S_\mathrm{synchro}(\nu)$ one thus obtains:
\begin{equation}
S_\mathrm{synchro}(\nu) = 
\sqrt{\frac{22~\mathrm{GHz}}{408~\mathrm{MHz}}}\cdot A_\mathrm{synchro}\cdot
\left(\frac{\nu}{408~\mathrm{MHz}}\right)^{-1.25}\mbox{.}
\end{equation}

Here, the fact that the synchrotron spectral index shows significant variations across the Milky Way due to varying magnetic 
field strength is ignored. Instead, a spatially constant spectral behaviour is assumed.

\subsection{Galactic free-free emission}\label{foreground_freefree}
The Galactic ionised plasma produces free-free emission, which is an important source of contamination in 
CMB observations, as recently confirmed by \citet{2003ApJS..148...97B} in WMAP observations. Aiming at modelling the 
free-free emission at microwave frequencies, we rely on an $H_\alpha$-template provided by \citet{2003ApJS..146..407F}. 
Modeling of the free-free emission component on the basis of an $H_alpha$-template is feasible because both emission 
processes depend on the emission measure $\int n_e^2\dd l$, where $n_e$ is the number density of electrons. This template 
is a composite of three $H_\alpha$-surveys and is because of its high resolution (on average $6\farcm0$ FWHM) particularly well 
suited for CMB foreground modelling. The morphology of the free-free map is very complex and the emission reaches out to 
intermediate Galactic latitude. 

For relating $H_\alpha$-fluxes $A_{H_\alpha}$ given in units of Rayleighs to the free-free signal's antenna 
temperature $T_\mathrm{free-free}$ measured in Kelvin, \citet{1998PASA...15..111V} gives the formula:
\begin{equation}
\frac{T_\mathrm{free-free}(\umu\mathrm{K})}{A_{H_\alpha}(R)} \simeq 
14.0\left(\frac{T_p}{10^4~\mathrm{K}}\right)^{0.317}\cdot 10^{290~\mathrm{K}\cdot T_p^{-1}}\cdot 
g_\mathrm{ff}\cdot\left(\frac{\nu}{10\mbox{ GHz}}\right)^{-2}\mbox{.}
\end{equation}
$T_p$ denotes the plasma temperature and is set to $10^4~\mathrm{K}$ in this work. An approximation for the 
Gaunt factor $g_\mathrm{ff}$ valid for microwave frequencies in the range $\nu_p\ll\nu\ll k_B T/h$ ($\nu_p$ is the 
plasma frequency) is given by \cite{2003ApJS..146..407F}:
\begin{equation}
g_\mathrm{ff} = \frac{\sqrt{3}}{\pi}
\left[\ln\left(\frac{(2 k_B T_p)^{3/2}}{\pi e^2 \nu\sqrt{m_e}}\right)-\frac{5}{2}\gamma_E\right]\mbox{,}
\end{equation}
where $e$ and $m_e$ denote electron charge and mass (in Gaussian units) and $\gamma_E\simeq0.57721$ is the Euler 
constant. The contribution of fractionally ionised helium to the free-free emissivity as well as the absorption by 
interstellar dust has been ignored because of its being only a small contribution in the first case and because of 
poorly known parameters in the latter case. The antenna temperature can be converted to the free-free flux 
$S_\mathrm{free-free}(\nu)$ by means of:
\begin{equation}
S_\mathrm{free-free}(\nu) = 2\frac{\nu^2}{c^2}\cdot k_B T_\mathrm{free-free}(\mathrm{K})\mbox{.}
\end{equation}

\begin{figure}
\resizebox{\hsize}{!}{\includegraphics{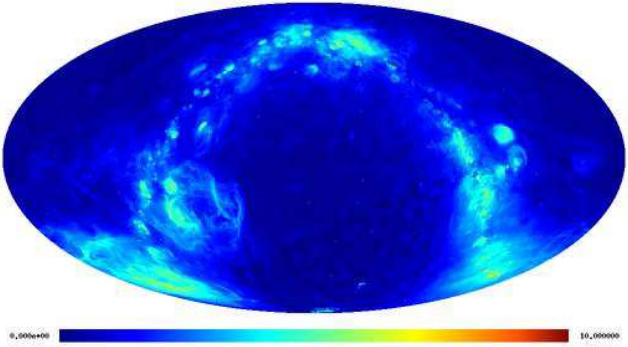}}
\caption{Sky map of the antenna temperature increase caused by free-free emission in the $\nu=100$~GHz-channel: The shading is 
proportional to $\mathrm{arsinh}(T_A(\nu=100~\mathrm{GHz}) / \mathrm{\umu K})$. Ecliptic coordinates have been chosen. This map 
has been derived from the $H_\alpha$-template map provided by \citet{2003ApJS..146..407F}.}
\label{figure_freefreemap}
\end{figure}

Concerning the free-free emission, there might be the possibility of an additional free-free component uncorrelated with the 
$H_\alpha$-emission. This hot gas, however, should emit X-ray line radiation, which has not been observed.

\subsection{CO-lines from giant molecular clouds}\label{foreground_co}
In a spiral galaxy such as the Milky Way, a large fraction of the interstellar medium is composed of molecular hydrogen, 
that resides in giant molecular clouds (GMC), objects with masses of $10^4 - 10^6 M_{\sun}$ and sizes of 
$50 - 200$~pc. Apart from molecular hydrogen, the GMCs contain carbon monoxide (CO) molecules in significant abundance. 
The rotational transitions of the CO molecule at 115~GHz and higher harmonics thereof constitute a source of 
contamination for all \planck~\hfi-channels. An extensive search for atomic and molecular transition lines was undertaken by 
\citet{1994ApJ...434..587B} with the {\em FIRAS} instrument onboard {\em COBE}.

\begin{figure}
\resizebox{\hsize}{!}{\includegraphics{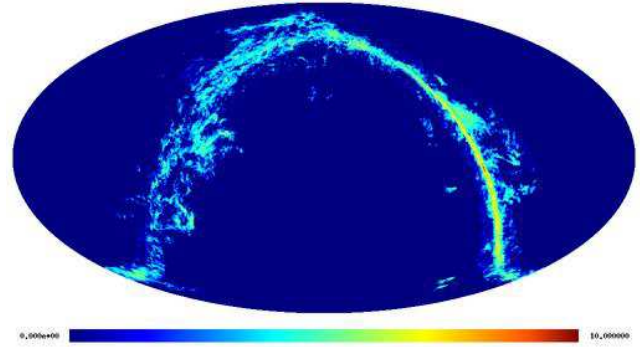}}
\caption{Sky map of the increment in antenna temperature due to CO-line emission in the $\nu=100$~GHz-channel in ecliptic 
coordinates: The shading is proportional to $\mathrm{arsinh}(T_A(\nu=100~\mathrm{GHz}) / \mathrm{\umu K})$. The maps shows the 
rotational transition of the CO molecule from the first excited state into the ground state at 
$\nu_{0\leftrightarrow1} = 115$~GHz as derived by \citet{2001ApJ...547..792D} for a temperature of $T_\mathrm{CO} = 20$~K.}
\label{figure_comap}
\end{figure}

The CO-contamination is modelled by employing a mosaic of CO-surveys assembled by \citet{1996AAS...189.7004D, 
2001ApJ...547..792D}. It shows the velocity-integrated intensity of the transition from the first excited state ($J=1$) to the 
ground state ($J=0)$ close to the Galactic plane ($b<5^\circ$), and additionally  comprises a few CO clouds at higher Galactic 
latitude, as well as the Large Magellanic Cloud and the Andromeda galaxy M~31. Due to the composition of the map, the angular 
resolution is not uniform, but the best resolution of $\simeq 7\farcm5$ is reached for a large area around the Galactic plane. 

From this map, it is possible to derive the line intensities of the higher harmonics, assuming thermal equilibrium:
The frequency $\nu$ for a transition from a state of rotational quantum number $J$ to a state with quantum number 
$J+1$ of the CO molecule follows from elementary quantum mechanics: The rotational energy of a molecule with moment of 
inertia $\theta$ and angular momentum $\bmath{J}$ is $E_\mathrm{rot}=\bmath{J}^2/2\theta=\hbar^2\cdot J(J+1)/2\theta$. In the 
last step the quantum number $J$ was introduced. For the transition energy between two subsequent rotation levels, one 
obtains:
\begin{equation}
\nu_{J\leftrightarrow J+1} = 2Qc\cdot(J+1) = 115\mbox{ GHz}\cdot(J+1)\mbox{,}
\end{equation}
where $Q=h/8\pi^2 c\theta$ is a measure of the inverse moment of inertia of the molecule and $c$ denotes the speed of light. 
Thus, the spectrum consists of equidistant lines. The relative intensities of those lines is given by the ratio of their 
occupation numbers $\chi_J$:
\begin{equation}\label{eqn_occupation}
\chi_J = (2J+1)\cdot\exp\left(-\frac{Qhc}{k_\mathrm{B} T_\mathrm{CO}} J(J+1)\right)\mbox{,}
\end{equation}
i.e. the relative line intensities $q_{J\leftrightarrow J+1}$ of two consecutive lines is given by:
\begin{equation}
q_{J\leftrightarrow J+1} = \frac{\chi_{J+1}}{\chi_J} = 
\frac{2J+3}{2J+1}\cdot\exp\left(-\frac{2Qhc}{k_B T_\mathrm{CO}}\cdot(J+1)\right)
\end{equation}
$\chi_J$ is detemined by a statistical weight $(2J+1)$ reflecting the degeneracy of angular momentum and a Boltzmann 
factor. For the determination of line intensities thermal equilibrium is assumed, common estimates for the temperature 
inside GMCs are $T_\mathrm{CO} = 10 - 30$~K. For the purpose of this work, we choose $T_\mathrm{CO} = 20$~K. From the 
brightness temperature $T_A$ one obtains the CO-flux $S_\mathrm{CO-line}(\nu)$ by means of the following equation:
\begin{equation}
S_\mathrm{CO-line}(\nu) = 2\frac{\nu^2}{c^2}\cdot k_B T_A(\mathrm{K})\cdot p(\nu-\nu_{J\leftrightarrow J+1})\mbox{,}
\end{equation}
where the line shape $p(\nu-\nu_{J\leftrightarrow J+1})$ is assumed to be small in comparison to \plancks frequency response 
windows such that its actual shape (for instance, a Voigt-profile) is irrelevant.

\subsection{Planetary submillimetric emission}\label{foreground_planets}
Planets produce infra-red and sub-millimetric radiation by absorbing sunlight and by re-emitting this thermal load 
imposed by the sun. The investigation of the thermal properties of Mars, Jupiter and Saturn has been the target of 
several space missions \citep[][ to name but a few]{1997ApJ...488L.161G, 1986Icar...65..244G}. For the description of the 
submillimetric thermal emission properties of planets, an extension to the Wright \& Odenwald model \citep{1976ApJ...210..250W, 
1971AJ.....76..719N} was used. The orbital motion of the planets is sufficiently fast such that their movements including their 
epicyclic motion relative to the Lagrangian point $L_2$, \plancks observing position, has to be taken into account. All planets 
are imaged twice in approximate half-year intervals due to \plancks scanning strategy, while showing tiny displacements from 
the ecliptic plane because of the Lissajous-orbit of \planck~around $L_2$ and their orbital inclinations.

The heat balance equation for a planet or asteroid reads as:
\begin{equation}
E + F + W \equiv P_\mathrm{emission} = P_\mathrm{absorption} \equiv I + R\mbox{,}
\end{equation}
where $E$ denotes the heat loss by thermal emission (i.e. the signal for \planck), $F$ the heat flux outward from 
the interior of the planet, $W$ is the heat lost by conduction to the planet's atmosphere, $I$ is the Solar radiation 
absorbed and $R$ is the heating of the planet caused by the back-scattering of radiation emanating 
from the surface of the planet by the atmosphere. The definition of these quantities is given by eqns.~(\ref{eqn_B_def}) 
through (\ref{eqn_R_def}):

\begin{eqnarray}
E & = & \epsilon\:\sigma\:T_\mathrm{planet}^4\mbox{,}\label{eqn_B_def}\\
F & = & k\cdot\frac{\upartial T_\mathrm{planet}}{\upartial x}\mbox{,}\label{eqn_F_def}\\
I & = & \frac{(1-A)G}{r^2}\cos(\theta^*)\cos\left(\frac{2\pi t}{\tau}\right)\mbox{,}\label{eqn_W_def}\\
R & = & \gamma\:\frac{(1-A)G}{r^2}\cos(\theta^*)\cos\left(\frac{2\pi t}{\tau}\right) = \gamma\:I_\mathrm{max}\mbox{, 
and}\label{eqn_I_def}\\
W & = & \kappa\:F\mbox{.}\label{eqn_R_def}
\end{eqnarray}
Here, $\epsilon$ is the surface emissivity of the planet, $\sigma$ is the Stefan-Boltzmann constant, $T_\mathrm{planet}$ is the 
planet's temperature, $k$ the coefficient of heat conduction, $A$ the planet's bolometric albedo, $G$ the Solar constant (i.e. 
the energy flux density of Solar irradiation at the earth's mean distance), $r$ the distance of the planet to the sun in 
astronomical units, $\tau$ the planet's rotation period and $\theta^*$ the geographical latitude of the radiation absorbing 
surface element. The temperature distribution in the interior of the planet at radial position $x$ is controlled by the heat 
conduction equation:
\begin{equation}
c\cdot\frac{\upartial T_\mathrm{planet}}{\upartial t} = k\cdot\frac{\upartial^2 T_\mathrm{planet}}{\upartial x^2}\mbox{,}
\end{equation}
with the specific heat per unit volume $c$.

In our model, the heat loss $R$ of the planet's surface due to conduction to the planet's atmosphere is 
taken to be a constant fraction of the heat flux $F$ outward from the interior of the planet, the constant of proportionality 
being $\kappa$, for which we assumed $\kappa=0.1$. Similarly, the heat gain by back-scattering radiation by the atmosphere $R$ 
was assumed to be a constant fraction $\gamma$ of the local noon Solar flux $I_\mathrm{max}$, where $\gamma$ was taken to be 
$\gamma=0.01$. The system of differential eqns.~(\ref{eqn_B_def}) - (\ref{eqn_R_def}) dependent on time $t$ and on Solar 
distance $r$ constitutes a heat conduction problem with periodic excitation (by the planet's rotation). Thus, the heat balance 
of the planets is modelled by periodic solutions of the Laplacian heat conduction differential equations. It was solved 
iteratively by applying Laplace transforms with periodic boundary conditions. The integration over the planet's surface then 
yields the radiation flux. In the calculation, we addressed rocky and gaseous planets differently with respect to their thermal 
properties. Furthermore, the giant gaseous planets are known to have  internal sources of heat generation, which also has been 
taken account of.

\begin{figure}
\resizebox{\hsize}{!}{\includegraphics{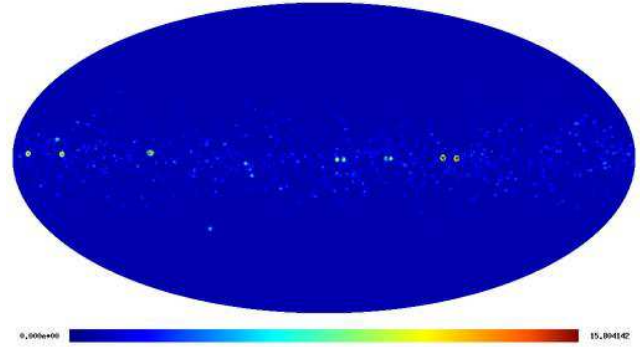}}
\caption{Sky map of increment in antenna temperature $T_A$ produced by planets and asteroids in the $\nu=30$~GHz channel at 
$33\farcm4$ resolution (FWHM). The colour coding is proportional to $\mathrm{arsinh}(T_A(\nu=100~\mathrm{GHz}) / 
n\mathrm{K})$. The asteroids reach ecliptic latitudes of $\left|\beta\right|\lsim30\degr$. The pronounced signals are 
produced by planets, which (due to \plancks scanning strategy) appear twice. The observable planets comprise (from left to 
right) Saturn, Mars, Uranus, Neptune and Jupiter. The epicyclic motion of Mars is sufficiently fast to counteract the 
parallactic displacement such that it appears only once.}
\label{figure_plamap}
\end{figure}

The brightest point source in the microwave sky due to the planetary thermal emission is Jupiter, causing 
an increase in antenna temperature of $T_\mathrm{Jupiter}=93.6$~mK in the $\nu=100$~GHz-channel, followed by Saturn with 
$T_\mathrm{Saturn}=15.0$~mK. All outer planets apart from Pluto will be visible for \planck. Estimates show that 
even Galilean satellites Ganymede, Callisto, Io and Europa and Saturn's moon Titan are above the detection threshold of  
\planck, but they are outshone by the stray-light from Jupiter and Saturn, respectively and for that reason not included in our 
analysis.

Due to the planet's being point sources, their fast movement and their diverse surface temperatures it is not feasible to 
produce a template and extrapolate the fluxes with a common emission law to \planck-frequencies. Instead, flux 
maps have been produced directly for each of the nine \planck-channels separately taking account of the planetary motion, the 
solution of the heat balance equation laid down above and the finite beam-width. The analogous holds for asteroids, that are 
covered by the next chapter.

\subsection{Submillimetric emission from asteroids}\label{foreground_asteroids}
Asteroids and minor bodies of the Solar system are easily observed by infrared satellites such as ISO and possibly by 
sub-millimetric observatories \citep{2001P&SS...49..787M, 1999DPS....31.0405M}. An estimation by 
\citet{2002NewA....7..483C} shows that a large number of asteroids ($\sim 400$) should yield signals detectable by 
\planck. The orbital motion of all asteroids is fast enough to cause double detections at different positions in the sky 
separated by half a year due to \plancks scanning strategy. In contrast to planets, asteroids are not well restricted to 
the ecliptic plane and appear up to ecliptic latitudes of $\beta\lsim 30\degr$.

The thermal emission properties of asteroids are well understood \citep[for a comprehensive and detailed review, 
see][]{1996A&A...310.1011L, 1996A&A...315..625L, 1997A&A...325.1226L, 1998A&A...332.1123L} such that asteroids have been 
used for calibrating detectors \citep[e.g. the ISO mission, c.f.][]{1998A&A...338..340M,
2002A&A...381..324M} and for determining beam shapes. The thermal model used for describing the 
submillimetric emission by asteroids is the same extension of the Wright \& Odenwald model as for rocky planets. 
However, additional features that had to be incorporated was the beamed emission due to surface roughness. 
Furthermore, in the system of differential eqns.~(\ref{eqn_B_def}) - (\ref{eqn_R_def}) terms $W$ and $R$ were neglected 
due to the absence of atmospheres in asteroids. 

Information about the diameter and albedo was derived using the HG-magnitude system in case of asteroids for which 
those quantities are unknown, otherwise literature values were taken \citep[from][ and IAU's {\em Minor Planet Centre} 
\footnote{\tt http://cfa-www.harvard.edu/cfa/ps/mpc.html}]{2000dba..book.....M}. For the description of the rotation 
period, an empirical relation that expresses the rotation period as a function of mass was used in the cases where the rotation 
period is unknown. The brightest sources include Ceres ($T_\mathrm{Ceres}=19.7~\umu\mbox{K}$), Pallas 
($T_\mathrm{Pallas}=7.2~\umu\mbox{K}$), Vesta ($T_\mathrm{Vesta}=6.7~\umu\mbox{K}$) and Davida 
($T_\mathrm{Davida}=2.1~\umu\mbox{K}$). The temperatures stated are antenna temperatures measured in the $\nu=100$~GHz-channel 
at the brightness maximum.

Our simulation shows that the number of detectable asteroids is overestimated by \citet{2002NewA....7..483C}, who did 
not take the expected observation geometry and detector response into account. Typical surface temperatures of asteroids 
are of the order of 150~K, and therefore, \planck~is observing their thermal emission in the Rayleigh-Jeans regime. 
For that reason, the number of detectable asteroids increases with observing frequency. For our sample of $5\cdot10^4$ 
asteroids of the {\em Minor Planet Centre}'s catalogue, we find a couple of asteroids at $\nu=30$~GHz, a few tens of 
asteroids at $\nu=100$~GHz and up to 100 asteroids in the highest frequency band at $\nu=857$~GHz. Approximately 1200 
asteroids will have fluxes above half of \plancks single-band detection limit estimated for ideal observation 
conditions and thus they constitute an abundant population of point sources that possibly hampers the detection of 
SZ-clusters. 

The prediction of comets is very uncertain for the years 2007 through 2009: Many comets are not detected yet, non-active 
comets are too faint with few exceptions and the coma thermal emission features of active comets is very complex. For 
these reasons, they have been excluded from the analysis.

\subsection{Future work concerning \plancks foregrounds}\label{foreground_omitted}
Foreground components not considered so far include microwave point sources, such as infra-red galaxies and 
microwave emitting AGNs. The emission of infra-red galaxies is associated with absorption of star light by dust and 
re-emission at longer wavelengths. Galaxies with ongoing star formation can have large fractions ($\sim90$\%) of their 
total emission at infra-red wavelengths, compared to about one third in the case of local galaxies. The integrated 
emission from unresolved infra-red galaxies accounts for the cosmic infra-red background (CIB) 
\citep{1996A&A...308L...5P, 2000A&A...355...17L}, the fluctuations of which are impeding SZ-observations at frequencies 
above $\nu\simeq100$~GHz \citep{2004astro.ph..2571A}. 

\citet{2003A&A...405..813L} and \citet{2003astro.ph..8464W} have estimated the number counts of unresolved infra-red 
galaxies at \planck~ frequencies, which was used by \citet{2004astro.ph..2571A} in order to estimate the level of 
fluctuation in the \planck-beam. In the easiest case, the sources are uncorrelated and the fluctuations obey 
Poissonian statistics, but the inclusion of correlations is expected to boost the fluctuations by a factor of $\sim1.7$ 
\citep{2003ApJ...590..664S}. According to \citet{2004astro.ph..2571A}, the resulting fluctuations vary between a few 
$10^2~\mathrm{Jy}/\mathrm{sr}$ and $10^5~\mathrm{Jy}/\mathrm{sr}$, depending on observing channel. A proper modeling would 
involve a biasing scheme for populating halos, the knowledge of the star formation history and template spectra in order 
to determine the K-corrections.

AGNs are another extragalactic source of submillimetric emission. Here, sychrotron emission is the radiation 
generating mechanism. The spectra show a variety of functional behaviours, with spectral indices $\alpha$ generally 
ranging from -1 to -0.5, but sources with inverted spectra $\alpha>0$ are commonplace. This variety makes it 
difficult to extrapolate fluxes to observing frequencies of CMB experiments. Two studies \citep{1998MNRAS.297..117T, 
2001ApJ...562...88S} have estimated the fluctuations generated by radio emitting AGNs at SZ-frequencies and found them 
to amount to $10^3-10^4~\mathrm{Jy}/\mathrm{sr}$. However, AGNs are known to reside in high-density environments and the 
proper modelling would involve a (poorly known) biasing scheme in order to assign AGN to the dark matter halos. Apart 
from that, one would have to assume spectral properties from a wide range of spectral indices and AGN activity duty 
cycles. Therefore, the study of extragalactic sources has been omitted from this analysis.

Yet another source of microwave emission in the Solar system is the zodiacal light \citep{2002A&A...393.1073L, 
2003astro.ph..4289R}. Modelling of this emission component is very difficult due to the Lissajous-orbit of 
\planck~around the Lagrangian point $L_2$. The disk of interplanetary dust is viewed under varying angles depending on 
the orbital period and the integration over the spatially non-uniform emission features is very complicated. 
\citet{2003Icar..164..384R} have investigated the thermal emission by interplanetary dust from measurements by ISO and 
have found dust temperatures of $T_\mathrm{zodiacal}=250-300$~K and fluxes on the level of 
$\simeq10^3~\mathrm{Jy}/\mathrm{sr}$, i.e. the equilibrium temperature is separated by two orders of magnitude from the 
CMB temperature, which means that the intensities are suppressed by a factor of $\sim10^4$ due to the Rayleigh-Jeans 
regime of the zodiacal emission in which \planck~is observing and by a factor of $10^5$ due to \plancks narrow 
beams. From this it is concluded that the emission from zodiacal light is unlikely to exceed values of a few 
$\sim\umu$Jy in observations by \planck~which compares to the fluxes generated by faint asteroids. Thus, the 
zodiacal light constitutes only a weak foreground emission component at submillimetric wavelengths and can safely be
neglected.

\section{Simulating SZ-observations by \planck}\label{sect_plancksim}
The simulation for assessing \plancks SZ-capabilities proceeds in four steps. Firstly, all-sky maps of the thermal 
and kinetic SZ-effects are prepared, the details of map-construction are given in Sect.~\ref{sim_szmap}. Secondly, a 
realisation of the CMB was prepared for the assumed cosmological model (Sect.~\ref{sim_cmbmap}). The amplitudes were co-added 
with the Galactic and ecliptic foregrounds introduced in the previous section, subsequently degraded in resolution with 
\plancks beams (Sect.~\ref{sim_beam}). Finally, uncorrelated pixel noise as well as the emission maps comprising planets and 
asteroids were added. In the last section, cross-correlation properties of the various astrophysical and instrumental noise 
components are discussed (Sect.~\ref{sim_ccproperties}).

At this stage it should be emphasised that we work exclusively with spherical harmonics expansion coefficients 
$a_{\ell m}$ of the flux maps. The expansion of a function $a(\bmath{\theta})$ into spherical harmonics 
$Y_\ell^m(\bmath{\theta})$ and the corresponding inversion is given by:
\begin{equation}
a_{\ell m} = \int\dd\Omega\: a(\bmath{\theta})\cdot Y_\ell^m(\bmath{\theta})^*\mbox{ and }
a(\bmath{\theta}) = \sum_{\ell=0}^{\infty}\sum_{m=-\ell}^{+\ell} a_{\ell m}\cdot Y_\ell^m(\bmath{\theta})\mbox{.}
\label{eqn_ylm_decomp}
\end{equation} 
Here, $\dd\Omega$ denotes the differential solid angle element. For reasons of computational feasibility, we assume isotropic 
spectral properties of each emission component, i.e. the template map is only providing the amplitude of the respective 
emission component, but the spectral dependences are assumed to remain the same throughout the sky. While this is an 
excellent approximation for the CMB and the SZ-effects (in the non-relativistic limit), it is a serious limitation for 
Galactic foregrounds, where e.g. the synchrotron spectral index or the dust temperatures show significant spatial 
variations. 

Adopting this approximation, the steps in constructing spherical harmonics expansion coefficients $\bra S_{\ell m}\ket_{\nu_0}$ 
of the flux maps $S(\bmath{\theta},\nu)$ for all \planck~channels consist of deriving the expansion coefficients of 
the template, converting the template amplitudes to flux units, extrapolate the fluxes with a known or assumed spectral 
emission law to \plancks observing frequencies, to finally convolve the emission law with \plancks~frequency response 
window for computing the spherical harmonics expansion coefficients of the average measured flux $\bra S_{\ell 
m}\ket_{\nu_0}$ at nominal frequency $\nu_0$ by using eqn.~(\ref{eqn_tlm_exp}).

\begin{equation}
\bra S_{\ell m}\ket_{\nu_0} 
= \frac{\int\dd\nu\: S_{\ell m}(\nu) R_{\nu_0}(\nu)}{\int\dd\nu\: R_{\nu_0}(\nu)} 
= 2\frac{\nu_0^2}{c^2}\cdot k_B T_{\ell m}\mbox{.}
\label{eqn_tlm_exp}
\end{equation}

Here, $S_{\ell m}(\nu)$ describes the spectral dependence of the emission component considered, and $R_{\nu_0}(\nu)$ the 
frequency response of \plancks receivers centered on the fiducial frequency $\nu_0$. Assuming spatial homogeneity of the 
spectral behaviour of each emission component it is possible to decompose $S_{\ell m}(\nu)$ into $S_{\ell m}(\nu) = q(\nu)\cdot 
a_{\ell m}$, i.e. a frequency dependent function $q(\nu)$ and the spherical harmonics expansion coefficients $a_{\ell m}$ of 
the template describing the morphology.  This is possible due to the fact that the decomposition eqn.~(\ref{eqn_ylm_decomp}) 
is linear. Additionally, eqn.~(\ref{eqn_tlm_exp}) gives the conversion from the averaged flux $\bra S_{\ell m}\ket_\nu$ in a 
\planck-channel to antenna temperature $T_{\ell m}$.

\plancks frequency response function $R_{\nu_0}(\nu)$ is well approximated by a top-hat function:
\begin{equation}
R_{\nu_0}(\nu) = 
\left\{
\begin{array}{l@{,\:}l}
1  & \nu\in\left[\nu_0-\Delta\nu,\nu_0+\Delta\nu\right] \\
0  & \nu\notin\left[\nu_0-\Delta\nu,\nu_0+\Delta\nu\right] 
\end{array}
\right.
\label{eq_freq_resp}
\end{equation}
The centre frequencies $\nu_0$ and frequency windows $\Delta\nu$ for \plancks receivers are summarised in 
Table.~\ref{table_planck_channel}. In this way it is possible to derive a channel-dependent prefactor relating the  
flux expansion coefficients $\bra S_{\ell m}\ket_{\nu_0}$ to the template expansion coefficients $A_{\ell m}$. The 
superposition of the various emission components in spherical harmonics and the determination of response-folded fluxes is most 
conveniently done using the {\tt almmixer}-utility of \plancks simulation package.

\subsection{SZ-map preparation}\label{sim_szmap}
For constructing an all-sky Sunyaev-Zel'dovich map, a hybrid approach has been pursued. Due to the SZ-clusters being 
detectable out to very large redshifts, due to their clustering properties on very large angular scales, and due to the 
requirement of reducing cosmic variance when simulating all-sky observations as will be performed by \planck, there 
is the need for very large simulation boxes, encompassing redshifts of $z\simeq1$ which corresponds to comoving scales 
exceeding 2~Gpc. Unfortunately, a simulation incorporating dark matter and gas dynamics that covers cosmological scales 
of that size down to cluster scales and possibly resolving cluster substructure is beyond computational feasibility. 
For that reason, two simulations have been combined: The Hubble-volume simulation \citep{2001MNRAS.321..372J, 
2000MNRAS.319..209C}, and a smaller scale simulation including (adiabatic) gas physics by \citet{2002ApJ...579...16W} 
performed with {\tt GADGET} \citep{2001NewA....6...79S, 2002MNRAS.333..649S}.

\begin{figure}
\resizebox{\hsize}{!}{\includegraphics{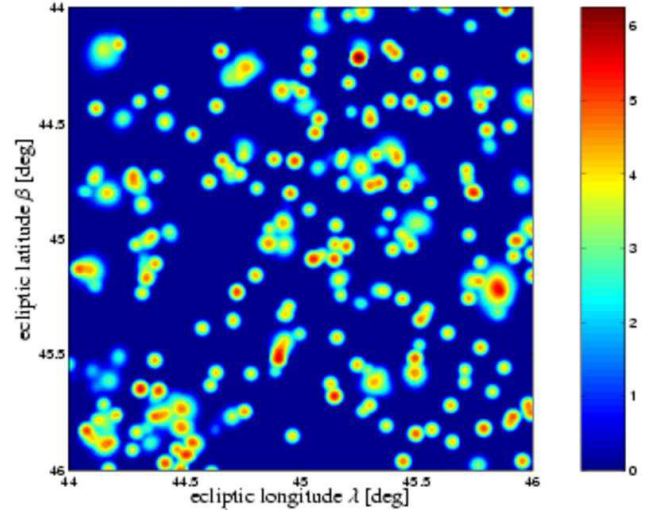}}
\caption{Detail of the thermal Comptonisation map: A $2\degr\times2\degr$ wide cut-out centered on the ecliptic 
coordinates $(\lambda,\beta) = (45\degr,45\degr)$ is shown. The smoothing imposed was a Gaussian kernel with 
$\Delta\theta = 2\farcm0$ (FWHM). The shading indicates the value of the thermal Comptonisation $y$, which is proportional to 
$\mathrm{arsinh}(10^6\cdot y)$. This map resulted from a projection on a Cartesian grid with mesh size $\sim14\arcsec$, i.e. 
no HEALPix pixelisation can be seen.}
\label{figure_thszfield}
\end{figure}

\begin{figure}
\resizebox{\hsize}{!}{\includegraphics{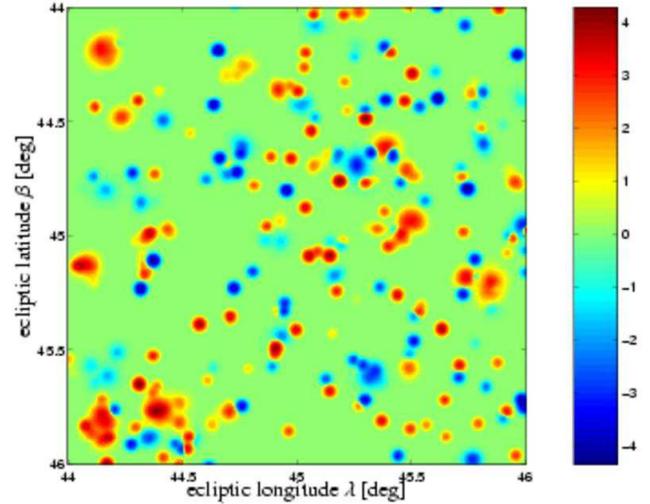}}
\caption{Detail of the kinetic Comptonisation map: A $2\degr\times2\degr$ wide cut-out centered on the same position as 
Fig.~\ref{figure_thszfield}, i.e. at the ecliptic coordinates $(\lambda,\beta) = (45\degr,45\degr)$ is shown. The 
smoothing imposed was a Gaussian kernel with $\Delta\theta = 2\farcm0$ (FWHM). The kinetic Comptonisation $w$ is 
indicated by the shading, being proportional to $\mathrm{arsinh}(10^6\cdot w)$.}
\label{figure_kinszfield}
\end{figure}

All-sky maps of the SZ-sky were constructed by using the light-cone output of the Hubble-volume simulation as a cluster 
catalogue and template clusters from the small-scale gas-dynamical simulation. In this way, the sky-maps contain all 
clusters above $5\cdot 10^{13} M_{\sun}/h$ out to redshift $z=1.48$. The analysis undertaken by 
\citet{2001A&A...370..754B} gives expected mass and redshift ranges for detectable thermal SZ-clusters, which are 
covered completely by the all-sky SZ-map presented here. The maps show the correct 2-point halo 
correlation function, incorporate the evolution of the mass function and the correct distribution of angular sizes.

Furthermore, they exhibit cluster substructure and deviations from the ideal cluster scaling relations induced by the 
departure from spherical symmetry. The velocities used for computing the kinetic SZ-effect correspond to the ambient 
density field. The map construction process and the properties of the resulting map are in detail described in 
\citet{2004_szmap}. Visual impressions of the SZ-maps are given by Figs.~\ref{figure_thszfield} and \ref{figure_kinszfield}.

The fluxes generated by the thermal SZ-effect $S_\mathcal{Y}(x)$ and of the kinetic SZ-effect $S_\mathcal{W}(x)$ are given 
by eqns.~(\ref{eq:S_thSZ}) and (\ref{eq:S_kinSZ}), respectively. The dimensionless frequency is defined as $x=h\nu/(k_B 
T_\mathrm{CMB})$ and the flux density of the CMB is given by 
$S_0=(k_b T_\mathrm{CMB})^3\pi^3/c^2/h^2/5400=22.9~\mathrm{Jy}/\mathrm{arcmin}^2$:
\begin{eqnarray}
S_\mathcal{Y}(x) & = & S_0\cdot\mathcal{Y}\cdot
\frac{x^4\cdot\exp(x)}{(\exp(x)-1)^2}\cdot\left[x\frac{\exp(x)+1}{\exp(x)-1} - 4\right]\mbox{.}
\label{eq:S_thSZ} \\
S_\mathcal{W}(x) & = & S_0\cdot\mathcal{W}\cdot\frac{x^4\cdot\exp(x)}{(\exp(x)-1)^2}\mbox{.}
\label{eq:S_kinSZ}
\end{eqnarray}

Table~\ref{table_planck_channel} summarises the fluxes $S_\mathcal{Y}$ and $S_\mathcal{W}$ and the corresponding changes 
in antenna temperature $T_\mathcal{Y}$ and $T_\mathcal{W}$ for the respective Comptonisation of 
$\mathcal{Y} = \mathcal{W} = 1~\mathrm{arcmin}^2$ for all \planck-channels.

\begin{figure}
\resizebox{\hsize}{!}{\includegraphics{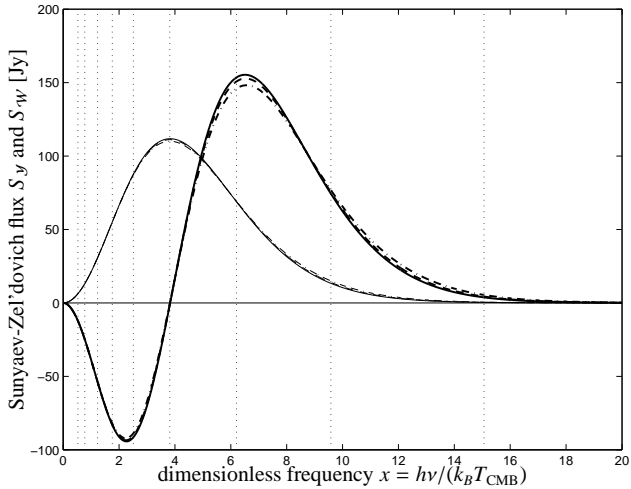}}
\caption{Frequency dependence of the thermal (thick lines) and of the kinetic SZ-flux (thin lines), for ideal $\delta$-like 
frequency responses (solid lines), for a top-hat window function with a relative width of 10\% corresponding to \plancks {\em 
LFI}-instrument and for a top-hat window function with a relative width of 16.7\%, as planned for \plancks {\em 
HFI}-instrument. The fluxes stated correspond to the integrated Comptonisation of 
$\mathcal{Y}=\mathcal{W}=1~\mathrm{arcmin}^2$. The vertical lines indicate the centre frequencies of \plancks receivers.}
\label{figure_sz_planck_window}
\end{figure}

Fig.~\ref{figure_sz_planck_window} shows how the frequency dependence of the SZ-signal is altered by \plancks relatively broad 
frequency response functions. The relative deviations of curves in which the frequency window has been taken into account to 
the unaltered curve amounts to $5\ldots15$\%, depending on observation frequency.

\subsection{CMB-map generation}\label{sim_cmbmap}
The angular power spectrum $C_\ell$ is computed for a flat \mbox{$\Lambda$CDM}-cosmology using the {\tt CMBfast}
code by \citet{1996ApJ...469..437S}. In addition to the cosmological parameters being already given in Sect.~\ref{sect_intro},
we use adiabatic initial conditions, set the CMB monopole to $T_\mathrm{CMB}=2.725$~K \citep{1999ApJ...512..511M} and 
the primordial He-mass fraction to $X_\mathrm{He} = 0.24$. The reionisation optical depth $\tau$ was set to $\tau=0.17$ and 
the reionisation redshift was taken to be $z_\mathrm{reion}=20$ \citep{2003ApJS..148....1B}. The angular power spectrum of the 
CMB is normalised to COBE data. With the spectrum of $C_\ell$-coefficients, a set of $a_{\ell m}$-coefficients was synthesised 
by using the {\tt synalm} code based on {\tt synfast} by \citet{1998elss.confE..47H}. The factors for converting the 
$a_{\ell m}$-coefficients of the CMB map showing the thermodynamic temperature and to the corresponding fluxes for each channel 
were then derived by convolution of the Planckian emission law eqn.~(\ref{eq:S_planck_cmb}),
\begin{equation}
S_\mathrm{CMB}(\nu) = S_0\cdot\frac{x^3}{\exp(x)-1}\mbox{,}
\label{eq:S_planck_cmb}
\end{equation}
with \plancks frequency response function eqns.~(\ref{eqn_tlm_exp}) and (\ref{eq_freq_resp}). Again,
$S_0=22.9~\mathrm{Jy}/\mathrm{arcmin}^2$ is the energy flux density of the CMB.

\subsection{Preparation of simulation data sets}\label{sim_beam}
The expansion coefficients of the flux maps are multiplied with the respective beam's $b_{\ell 0}$-coefficients 
in order to describe the finite angular resolution. After that, expansion coefficients of the pixel noise maps and those of 
the planetary maps have been added. In total, three atlases consisting of nine flux $\bra S_{\ell m}\ket_{\nu_0}$-sets 
belonging to each of \plancks channels with fiducial frequency $\nu_0$ have been compiled:
\begin{itemize}
\item{The reference data set is a combination of the CMB, the SZ-maps and the instrumental noise maps. They should 
provide the cleanest detection of clusters and the measurement of their properties. Apart from the inevitable 
instrumental noise, this data set only contains cosmological components. In the remainder of the paper, this data set 
will be refered to as {\tt COS}.}

\item{The second data set adds Galactic foregrounds to the CMB, the SZ-maps and the instrumental noise map. 
Here, we try to assess the extend to which Galactic foregrounds impede the SZ-observations. Thus, this data set will be 
denoted {\tt GAL}.}

\item{In the third data set the emission from bodies inside the Solar system was included to the CMB, the SZ-maps, the 
Galactic foregrounds and the instrumental noise. Because of the planets and asteroids being loosely constrained to the ecliptic 
plane, this data set will be called {\tt ECL}.}
\end{itemize}

\begin{figure}
\resizebox{\hsize}{!}{\includegraphics{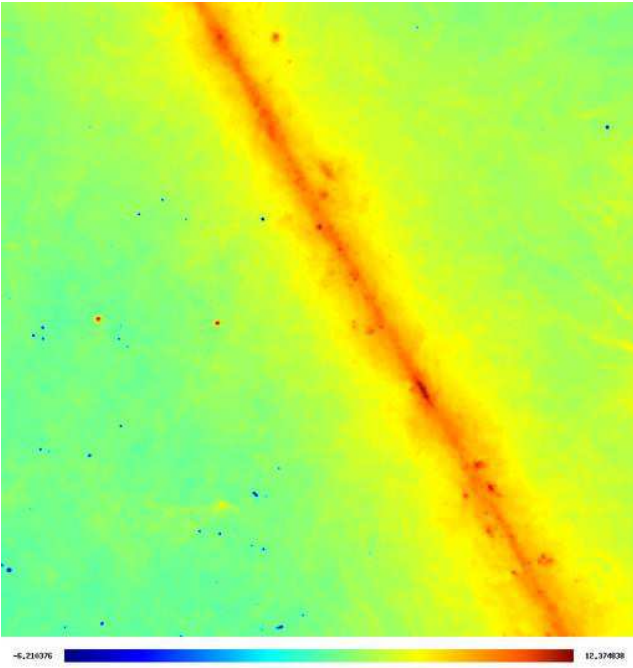}}
\caption{A $50\degr\times50\degr$ wide composite centered on the ecliptic coordinates $(\lambda,\beta) = (-85\degr,0\degr)$, 
i.e. close to the Galactic centre for \plancks $\nu=143$~GHz-channel. The shading is proportional to 
$\mathrm{arsinh}(T_A(\nu=143~\mathrm{GHz})/\mathrm{\umu K})$. The map is smoothed with the corresponding beam of diameter 
$\Delta\theta = 7\farcm1$ (FWHM). SZ-clusters are observed in absorption in this channel and are discernible by eye even at 
close proximity ($b\lsim20\degr$) to the Galactic plane. For clarity, the CMB fluctuations as well as the instrumental noise 
have been excluded. The two point sources on the ecliptic equator are twin detections of Jupiter.}
\label{figure_sim_skyview}
\end{figure}

An example of a synthesised map showing the combined emission of the SZ-clusters and all Galactic and ecliptic components  
including neither CMB fluctuations nor instrumental noise at a location close to the Galactic plane is given by 
Fig.~\ref{figure_sim_skyview}. The observing frequency has been chosen to be $\nu = 143$~GHz, correspondingly, the map has been 
smoothed with a (Gaussian) beam of $\Delta\theta=7\farcm1$ (FWHM).

\subsection{\planck-channel correlation properties}\label{sim_ccproperties}
\begin{figure}
\resizebox{\hsize}{!}{\includegraphics{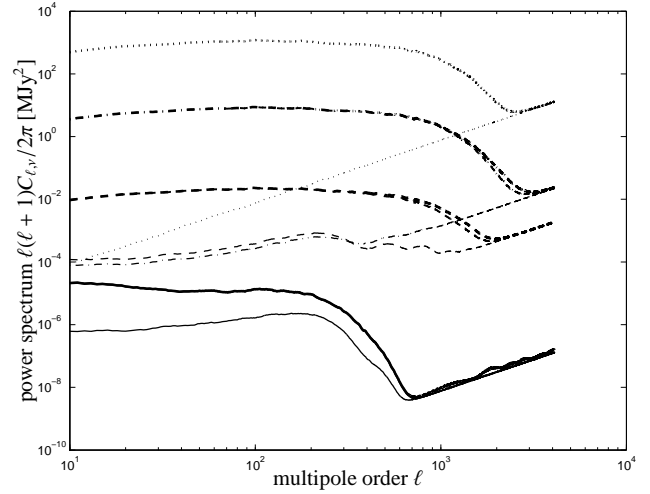}}
\caption{Power spectra in various \planck-channels: $\nu=30$~GHz (solid), $\nu=143$~GHz (dashed), 
$\nu=353$~GHz (dash-dotted) and $\nu=857$~GHz (dotted) for {\tt COS} data set (thin line), the {\tt GAL} data set (medium 
line) and the {\tt ECL} data set (thick line).}
\label{figure_auto_correlation}
\end{figure}

In this section the auto- as well as the cross-correlation properties of the various foregrounds in different \planck-channels 
are studied. The cross power specta, defined formally by eqn.~(\ref{eqn_cross_power_def}) are determined by using:
\begin{equation}
C_{\ell,\nu_1\nu_2} = \frac{1}{2\ell+1}\sum_{m=-\ell}^{+\ell} 
\bra S_{\ell m}\ket_{\nu_1}\cdot \bra S_{\ell m}\ket_{\nu_2}^*\mbox{.}
\label{eqn_cross_power}
\end{equation}
From this definition, the auto-correlation spectra are obtained by setting $\nu_1=\nu_2$, i.e. $C_{\ell,\nu} = 
C_{\ell,\nu\nu}$. The band-pass averaged fluxes $\bra S_{\ell m}\ket_\nu$ are defined in eqn.~(\ref{eqn_tlm_exp}). In 
Fig.~\ref{figure_auto_correlation}, the power spectra are shown for the $\nu=30$~GHz-, $\nu=143$~GHz-, $\nu=353$~GHz- and the 
$\nu=847$~GHz-channels. The spectra have been derived including various Galactic and ecliptic noise components in order to 
study their relative influences. For visualisation purposes, the spectra are smoothed with a moving average filter with a 
filter window comprising 11 bins.

Distinct acoustic peaks of the CMB are clearly visible in the clean {\tt COS} data sets, but are overwhelmed by the Galactic 
noise components. At small scales, i.e. high multipole order $\ell$, differences between the {\tt GAL} and {\tt ECL} data sets 
become apparent, the latter showing a higher amplitude. The (single) acoustic peak measurable in the $\nu=33$~GHz channel is 
shifted to larger angular scales due to the coarse angular resolution of that particular channel. The $\nu=857$~GHz-curve of 
the {\tt COS} data set behaves like a power law due to the fact that the CMB is observed in the Wien-regime and is consequently 
strongly suppressed, such that the angular power spectrum is dominated by uncorrelated pixel noise. 

\begin{figure}
\resizebox{\hsize}{!}{\includegraphics{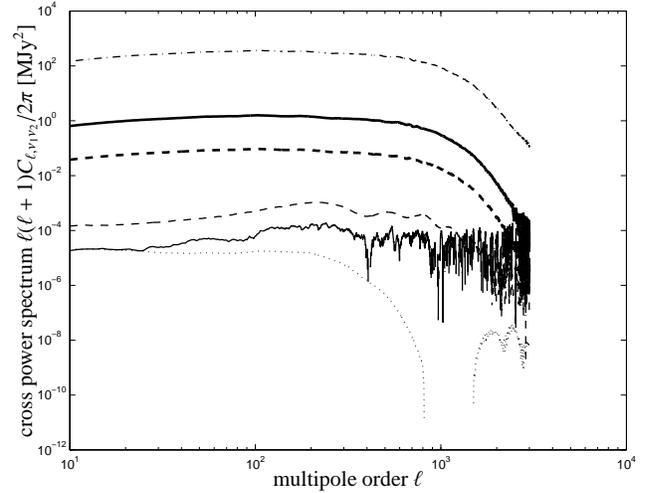}}
\caption{Cross-correlations: The spectra $C_{\ell,\nu_1=143~\mathrm{GHz},\nu_2=545~\mathrm{GHz}}$ (solid line) and 
$C_{\ell,\nu_1=143~\mathrm{GHz},\nu_2=217~\mathrm{GHz}}$ (dashed line) are contrasted for the {\tt COS} data set (thin lines) 
versus the {\tt GAL} data set (thick line). Furthermore, the spectrum $C_{\ell,\nu_1=545~\mathrm{GHz},\nu_2=857~\mathrm{GHz}}$ 
(dash-dotted line) as well as $C_{\ell,\nu_1=30~\mathrm{GHz},\nu_2=44~\mathrm{GHz}}$ (dotted line) is shown as derived from the 
{\tt ECL} data set.}
\label{figure_cross_correlation}
\end{figure}

Fig.~\ref{figure_cross_correlation} shows exemplarily a couple of cross power spectra. The cross-correlation spectra derived 
for the {\tt COS} data set nicely shows the CMB power spectrum if two neighboring channels close to the CMB maximum are chosen, 
but the correlation is lost in two widely separated channels. This is especially the case if one considers the two 
lowest {\em LFI}-channels at angular scales which the receivers are not able to resolve. In this regime the pixel noise is 
still very small and the cross-correlation spectrum drops to very small values. 

\begin{figure}
\resizebox{\hsize}{!}{\includegraphics{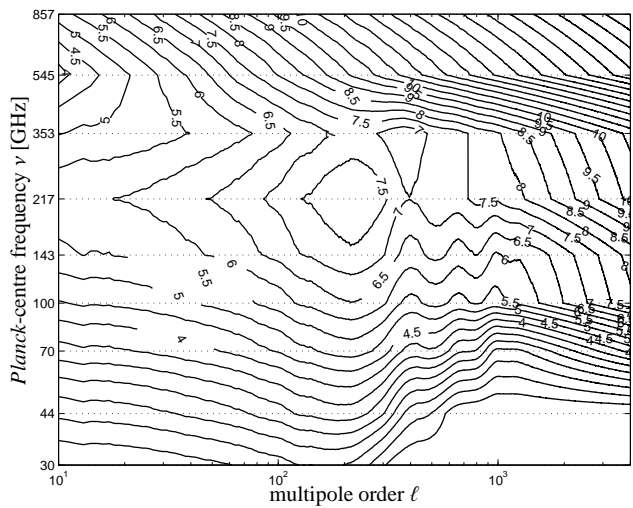}}
\caption{Auto-correlations: The power spectral $C_{\ell,\nu}$-coefficients are shown as 
a function of observing frequency $\nu$ and multipole order $\ell$ in the usual representation $\ell(\ell+1)C_{\ell,\nu}/2\pi$. 
The amplitudes are given in $\mathrm{\umu K}^2$ and the contours are linearly spaced. Note the logarithmic scaling of the 
frequency axis. In the data set displayed, the CMB, both SZ-effects and instrumental noise are included. The first three 
acoustic oscillation peaks are clearly visible.}
\label{figure_cos_correlation_contour}
\end{figure}

\begin{figure}
\resizebox{\hsize}{!}{\includegraphics{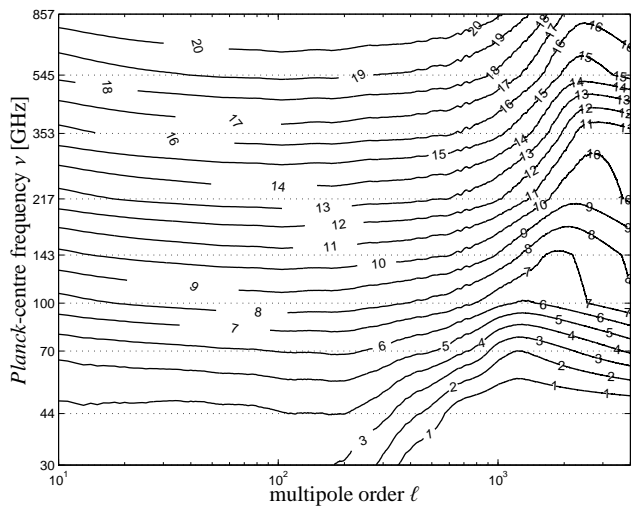}}
\caption{Auto-correlations: The power spectral $C_{\ell,\nu}$-coefficients are shown as 
a function of observing frequency $\nu$ and multipole order $\ell$ in the usual representation $\ell(\ell+1)C_{\ell,\nu}/2\pi$. 
The amplitudes are given in $\log(\mathrm{\umu K}^2)$ with logarithmically equidistant contours. In the data set 
displayed, the CMB, both SZ-effects, all Galactic foregrounds and instrumental noise are included.}
\label{figure_gal_correlation_contour}
\end{figure}

In order to illustrate the complexity of spectral and morphological behaviour of the power spectra, they are given as contour 
plots depending on both the observing frequency $\nu$ and the multipole order $\ell$. 
Fig.~\ref{figure_cos_correlation_contour} and \ref{figure_gal_correlation_contour} contrast the auto-correlation properties of 
the different data sets. The {\tt COS} data set, shown in Fig.~\ref{figure_cos_correlation_contour}, containing nothing but the 
CMB and instrumental noise apart from the SZ-contribution, shows clearly the acoustic oscillations with the first peak at 
$\ell\simeq200$ and the consecutive higher harmonics. They are most pronouced in the $\nu=100$~GHz- and $\nu=143$~GHz-channels. 
At higher multipole moments, the power spectra are dominated by instrumental noise which leads to a rapid (power law) incline.

Adding Galactic foregrounds yields the spectra depicted in Fig.~\ref{figure_gal_correlation_contour}. Inclusion of Galactic 
foregrounds significantly complicates the picture and masks off the primary anisotropies. The spectra are dominated by 
large-scale emission structures of the Milky Way, most notably the emission from thermal dust that causes the spectra to 
increase with increasing frequency $\nu$.

\section{Cluster detection by using multi-frequency optimised filtering}\label{sect_filtering}
One challenge in the analysis of two-dimensional all-sky surveys is the extraction of sources of interest which are 
superposed on a background of noise of varying morphology and spectral behaviour.  In the presence of small-scale noise the 
conventional method to extract sources is low-pass filtering (e.g.~with a Gaussian kernel) while wavelet analysis is most 
suitably applied if large scale noise fluctuations dominate. These methods, however, fail if the characteristic scale of the 
background fluctuations is comparable with the scale of the signal structures. Other methods have been proposed in
order to separate different components in multifrequency CMB observations: They include Wiener filtering 
\citep{1996MNRAS.281.1297T, 1999NewA....4..443B,  1999MNRAS.302..663B}, maximum-entropy methods \citep{1998MNRAS.300....1H, 
1999MNRAS.306..232H}, Mexican-hat wavelet analysis \citep{2001MNRAS.326..181V, 2000MNRAS.315..757C}, fast independent 
component analysis \citep{2002MNRAS.334...53M}, matched filter analysis \citep{1998ApJ...500L..83T}, adaptive filtering 
techniques \citep{2001ApJ...552..484S, 2002ApJ...580..610H, 2002MNRAS.336.1057H}, and non-parametric Bayesian approaches 
\citep{2002MNRAS.336.1351D}.

However, a comparison between these methods is difficult because all of them assume different priors about the spatial 
properties and frequency dependence. Using prior knowledge about the frequency dependence and statistical properties
of several images at different frequency channels, the maximum-entropy method and Wiener filtering are able to separate the 
components of interest. Contrarily, wavelet analysis is well suited in order to detect compact sources. A combination of these 
different techniques improves the quality of component separation \citep{2001MNRAS.328....1V}. Although component separation 
methods which assume a prior knowledge about the data are quite powerful, they yield biased or even wrong results in the case 
of incorrect or idealised assumptions about the data. Any error in the separation of one component propagates to the 
separation of the other components owing to normalisation constraints.  In particular, this is the case in non-centrally 
symmetric source profiles, oversimplified spectral extrapolations of Galactic emission surveys into other wavebands, variations 
of the assumed frequency dependence, or non-Gaussian noise properties the statistics of which can not fully be characterised by 
power spectra. Thus, the application of a specific component separation method is a trade-off between robustness and 
effectiveness with regard to the particular problem.

Filtering techniques relying on Mexican-hat wavelets and on matched and scale-adaptive filters are single component separation 
methods. They all project either spatial structure or frequency properties (within a given functional family) of the component 
of interest in the presence of other components acting as background in this context. While Mexican-hat wavelet analysis 
assumes Gaussian profiles superimposed on large scale variations of the background noise, the matched and scale-adaptive filter 
generalises to arbitrary source profiles and noise properties which are assumed to be locally homogeneous and isotropic 
\citep{2001ApJ...552..484S, 2002MNRAS.336.1057H, 2002ApJ...580..610H}.
  
This section generalises the matched and scale-adaptive filter techniques to global spherical topologies which find 
application in all-sky surveys such as the case of \plancks microwave/submillimetric survey. In addition, optimised filters 
for the detection of compact sources in single frequency all-sky observations are derived in the appendix in a more detailed 
fashion. The proposed method aims at simultaneously localising SZ-clusters and measuring both their amplitudes and angular 
extent. It can also be applied for localising microwave point sources and estimating their spectral properties.

We choose the spherical filtering approach rather than tiling the sky with a set of two-dimensional flat maps for the 
following reasons: On the sphere, we do not have to worry about spurious detections as well as double detections due to 
overlaps in the tesselation. Secondly, our approach provides a physical interpretation of our filter shapes in harmonic space 
even for the smallest multipole moments in contrast to the case of a flat map where the smallest wavenumbers are determined by 
the map size. Finally, our approach circumvents projection failures of the noise properties such as stretching effects in the 
case of conformal mapping which would introduce artifical non-Gaussianity in our maps and distort profile shapes close to the 
map boundaries.

We pursue the concept of the {\em multi-frequency approach} rather than the {\em combination method} 
\citep[c.f.][]{2002MNRAS.336.1057H}. In other words, we filter each channel separately while taking into account the different 
cross-correlations between the different channels and the frequency dependence of the signal when constructing the optimised 
filters.  This method seems to be superior to the {\em combination  method} which tries to find a optimised combination of the 
different channels with regard to the signal-to-noise ratio of the sources and successively applies filters to the combined 
map.

The concept is introduced and central definitions are laid down in Sect.~\ref{filter_construct}. The concept of 
constructing filter kernels is outlined in Sect.~\ref{filter_optimal}. Subsequently, the matched and scale-adaptive filters 
are derived for expansions of spherical data sets into spherical harmonics in Sect.~\ref{filter_allsky_matched} and 
Sect.~\ref{filter_allsky_scaleadaptive}. Then, the numbers of merit are defined in Sect.~\ref{filter_gain_reliability}. 
Caveats in the numerical derivation are listed in Sect.~\ref{filter_numerics}. A discussion of filter kernel shapes in 
Sect.~\ref{filter_shape_discussion} for actual simulation data. The application of the filter kernels to our simulated sky maps 
and the extraction of the SZ-cluster signal is described in Sect.~\ref{filter_renormalise}.

\subsection{Assumptions and definitions}\label{filter_construct}
When constructing the particular filters, we assume centrally symmetric profiles of the sources to be detected. This 
approximation is justified for most of the clusters of \plancks sample whose angular extent will be comparable in size to 
\plancks beams, i.e. the instrumental beam renders them azimuthally symmetric irrespective of their intrinsic shape. 
Azimuthal symmetry is no general requirement for the filters which can be generalised to detect e.g. elliptic clusters using 
expansions into vector rather than scalar spherical harmonics. 

We furthermore assume the background to be statistically homogeneous and isotropic, i.e.{\ }a complete characterisation can be 
given in terms of the power spectrum. This assumption obviously fails for non-Gaussian emission features of the Galaxy or of 
the exposure-weighted instrumental noise on large angular scales. However, the spherical harmonics expansion of any expected 
compact source profile, which we aim to separate, peaks at high values of the multipole moment due to the smallness of the 
clusters where the non-Gaussian influence is negligible. Thus, we only have to require homogeneity and isotropy of 
the background on small scales.

In order to construct our filters, we consider a set of all-sky maps of the detected scalar field $s_\nu(\btheta)$ for the 
different frequency channels
\begin{equation}
  \label{eq:signal}
  s_\nu(\btheta) = f_\nu y_\nu(|\btheta-\btheta_0|) + n_\nu(\btheta), 
  \quad \nu = 1, \ldots, N,
\end{equation}
where $\btheta = (\vartheta, \varphi)$ denotes a two-dimensional vector on the sphere, $\btheta_0$ is the source location, and 
$N$ is the number of frequencies (respectively, the number of maps).  The first term on the right-hand side represents the 
amplitude of the signal caused by the thermal and kinetic SZ-effect, $y(|\btheta-\btheta_0|)$ and $w(|\btheta-\btheta_0|)$, 
respectively, while the second term corresponds to the generalised noise which is composed of CMB radiation, all Galactic and 
ecliptic emission components, and additional instrumental noise.  The frequency dependence of the SZ-effect is described by 
$f_\nu$ in terms of average flux,
\begin{equation}
  \label{eq:fnu}
  f_\nu \equiv \bra S_\mathcal{Y}\ket_{\nu}\mbox{ and } f_\nu \equiv \bra S_\mathcal{W}\ket_{\nu}
\end{equation}
where $\bra S\ket_\nu$ denotes the flux weighted by the frequency response at 
the fiducial frequency $\nu$ (c.f. eqn.~(\ref{eqn_tlm_exp})) and $S_\mathcal{Y}$ and $S_\mathcal{W}$ denote the SZ-fluxes 
given by eqns.~(\ref{eq:S_thSZ}) and (\ref{eq:S_kinSZ}). 

We expect a multitude of clusters to be present in our all-sky maps. In order to sketch the construction of the optimised 
filter, we assume an individual cluster situated at the North pole ($\btheta_0=\bld{0}$) with a characteristic angular 
SZ-signal $y_\nu(\theta = |\btheta|) = A \tau_\nu(\theta)$, where we separate the true amplitude $A$ and the spatial profile 
normalised to unity, $\tau_\nu(\theta)$. The underlying cluster profile $p(\theta)$ is assumed to follow a generalised 
King-profile with an exponent $\lambda$ which is a parameter in our analysis. At each observation frequency this profile is 
convolved with the (Gaussian) beam of the respective \planck-channel (c.f.~Sect.~\ref{sect_beam}) yielding:
\begin{eqnarray}
  \label{eq:profile}
  \tau_\nu^{}(\theta) &=& 
  \int \dd\Omega' p(\theta') b_\nu^{}(|\btheta-\btheta'|) 
  =\sum_{\ell=0}^{\infty}\tau_{\ell 0,\, \nu}^{}  Y_\ell^0(\cos \theta),\\
  p(\theta) &=& 
  \left[1+\left(\frac{\theta}{\theta_\rmn{c}}\right)^2\right]^{-\lambda}\!\!, 
  \quad\mbox{and}\quad \tau_{\ell 0,\, \nu} 
  = \sqrt{\frac{4 \pi}{2 \ell+1}} b_{\ell 0,\, \nu} p_{\ell 0}.
  \label{eqn_profile_beam_source}
\end{eqnarray}
For the second step in eqn.~(\ref{eqn_profile_beam_source}) we used the convolution theorem on the sphere to be derived in 
Appendix~\ref{appendix_sphsingle}. The background $n_{\nu}(\btheta)$ is assumed to be a compensated homogeneous and isotropic 
random field with a cross power spectrum $C_{\ell, \nu_1 \nu_2}$ defined by
\begin{equation}
  \label{eq:PSnoise}
  \left\bra n_{\ell m, \nu_1}^{} n^*_{\ell' m', \nu_2} \right\ket =
  C_{\ell, \nu_1 \nu_2}^{} \delta_{\ell \ell'}^{} \delta_{m,m'}^{},
  \quad\mbox{where}\quad \bra n_{\nu}(\btheta) \ket = 0,
\label{eqn_cross_power_def}
\end{equation}
$n_{\ell m, \nu}$ denotes the spherical harmonics expansion coefficient of $n_{\nu}(\btheta)$, $\delta_{\ell \ell'}$ denotes 
the Kronecker symbol, and $\bra\cdot\ket$ corresponds to an ensemble average.  Assuming ergodicity of the field under 
consideration allows taking spatial averages over sufficiently large areas $\Omega = \mathcal{O}(4\pi) $ instead of performing 
the ensemble average.

\subsection{Concepts in filter construction}\label{filter_optimal}
The idea of an optimised matched filter for multifrequency observations was recently proposed by \citet{2002MNRAS.336.1057H} 
for the case of a flat geometry. For each observing frequency, we aim at constructing a centrally symmetric optimised filter 
function $\psi_\nu(\theta)$ operating on a sphere. Its functional behaviour induces a family of filters 
$\psi_\nu(\theta,R_\nu)$ which differ only by a scaling parameter $R_\nu$. For a particular choice of this parameter, we define 
the filtered field $u_\nu(R_\nu,\bbeta)$ to be the convolution of the filter function with the observed all-sky map at 
frequency $\nu$,
\begin{eqnarray}
\label{eq:udef}
u_\nu^{}(R_\nu,\bbeta) & = & \int \dd\Omega\, s_\nu^{}(\btheta)\,\psi_\nu^{}(|\btheta-\bbeta|,R_\nu) \\
                       & = & \sum_{\ell=0}^{\infty}\sum_{m=-\ell}^{+\ell} u_{\ell m,\, \nu}^{}  Y_\ell^m(\bbeta)\mbox{ with}\\
u_{\ell m,\, \nu}      & = & 
\sqrt{\frac{4 \pi}{2 \ell+1}} s_{\ell m,\, \nu}\, \psi_{\ell 0,\, \nu}(R_\nu) \,\label{eqn_k_convolve}.
\end{eqnarray}

For the second step, the convolution theorem to be derived in Appendix~\ref{appendix_sphsingle} was used. The combined 
filtered field is defined by
\begin{equation}
  \label{eq:utotal}
  u(R_1,\ldots,R_N;\bbeta)=\sum_{\nu} u_\nu(R_\nu,\bbeta).
  \label{eqn_k_add}
\end{equation}
Taking into account the vanishing expectation value of the noise $\bra n_{\nu}(\btheta) \ket = 0$, the expectation value of the 
filtered field at the North pole $\bbeta=\bld{0}$ is given by
\begin{equation}
  \label{eq:umean}
  \bra u_\nu(R_\nu,\bld{0})\ket = 
  A f_\nu \sum_{\ell=0}^{\infty} \tau_{\ell 0,\, \nu}\,
  \psi_{\ell 0,\, \nu}(R_\nu).
  \label{eqn_def_ccspec}
\end{equation}
The assumption that the cross power spectrum of the signal is negligible compared to the noise power spectrum is justified 
because the thermal and kinetic amplitudes are small compared to unity, $A_{y,w} \ll 1$. Thus, the variance of the combined 
filtered field (\ref{eq:utotal}) is determined by
\begin{eqnarray}
  \label{eq:uvariance}
  \sigma_u^2(R_1,\ldots,R_N) &=&
  \left\bra \left[u(R_1,\ldots,R_N;\bbeta) - 
    \left\bra u(R_1,\ldots,R_N;\bbeta)\right\ket\right]^2\right\ket 
  \nonumber\\   
  &=&\sum_{\nu_1,\nu_2}\sum_{\ell=0}^{\infty} C_{\ell,\,\nu_1 \nu_2}
  \psi_{\ell 0,\, \nu_1}(R_{\nu_1})\,\psi_{\ell 0,\, \nu_2}(R_{\nu_2}).
\end{eqnarray}

The optimised filter functions $\psi_\nu(\theta)$ are chosen to detect the clusters at the North pole of the sphere (to which 
they have been translated). They are described by a singly peaked profile which is characterised by the scale $R^{(0)}_{\nu}$ 
as given by eqn.~(\ref{eq:profile}).  While the optimised {\em matched filter} is defined to obey the first two of the 
following conditions, the optimised {\em scale-adaptive filter} is required to obey all three conditions:

\begin{enumerate}
\item{The combined filtered field $u(R^{(0)}_1,\ldots,R^{(0)}_N;\bld{0})$ is an unbiased estimator of the source amplitude 
$A$, i.e.{\ }$\bra  u(R^{(0)}_{1},\ldots,R^{(0)}_{N};\bld{0})\ket = A$.}
  
\item{The variance of $u(R_1,\ldots,R_N;\bbeta)$ has a minimum at the scales $R^{(0)}_1,\ldots,R^{(0)}_N$ ensuring that the 
combined filtered  field is an efficient estimator.}
  
\item{The expectation value of the filtered field at the source position has an extremum with respect to the the scale 
$R^{(0)}_{\nu}$, implying  
\begin{equation}
\label{eq:3cond}
\frac{\upartial}{\upartial R^{(0)}_{\nu}}\bra u_\nu(R_\nu,\bld{0})\ket = 0.
\end{equation}}
\end{enumerate}

\subsection{Matched filter}\label{filter_allsky_matched}
For convenience, we introduce the column vectors $\bpsi_{\ell} \equiv [\psi_{\ell 0,\,\nu}]$, $\bld{F}_{\ell} \equiv [f_\nu 
\tau_{\ell 0, \nu}]$, and the inverse $\hat{\bld{C}}_{\ell}^{-1}$ of the matrix $\hat{\bld{C}}_{\ell} \equiv [C_{\ell,\,\nu_1 
\nu_2}]$.  In terms of spherical harmonic expansion coefficients, constraint (i) reads
\begin{equation}
  \label{eq:constraint1}
  \sum_\nu \sum_{\ell=0}^\infty 
  f_\nu \tau_{\ell 0,\, \nu} \psi_{\ell 0,\, \nu} = 
  \sum_{\ell=0}^\infty \bld{F}_{\ell}\bpsi_{\ell} = 1. 
\end{equation}
Performing functional variation (with respect to the filter function $\bpsi_{\ell}$) of $\sigma_u^2(R_1,\ldots,R_N)$ while 
incorporating the (isoperimetric) boundary condition (\ref{eq:constraint1}) through a Lagrangian multiplier yields the
spherical matched filter $\bpsi_{\ell}$
\begin{equation}
  \label{eq:matched filter}
  \bpsi_{\ell}^{} = \alpha\, \hat{\bld{C}}_{\ell}^{-1} \bld{F}_{\ell}^{},
  \quad\mbox{where}\quad  \alpha^{-1} = \sum_{\ell=0}^\infty 
  \bld{F}_{\ell}^T \hat{\bld{C}}_{\ell}^{-1} \bld{F}_{\ell}^{}.
\end{equation}
In any realistic application, the cross power spectrum $C_{\ell, \nu_1 \nu_2}$ can be computed from observed data provided 
the cross power spectrum of the signal is negligible. The quantities $\alpha$,  $\bld{F}_{\ell 0}$, and thus $\bpsi_{\ell 0}$ 
can be computed in a straightforward manner for a specific frequency dependence $f_\nu$ and for a model source profile 
$\tau_\nu(\theta)$.

\subsection{Scale-adaptive filter on the sphere}\label{filter_allsky_scaleadaptive}
The scale-adaptive filter $\bpsi_{\ell}$ satisfying all three conditions is given by
\begin{equation}
  \label{eq:SAF}
  \bpsi_{\ell}^{} = \hat{\bld{C}}_{\ell}^{-1} (\alpha\, \bld{F}_{\ell}^{}+\bld{G}_{\ell}^{})
\mbox{, with } \bld{G}_{\ell}^{} \equiv [\mu_{\ell,\nu}^{}\, \beta_\nu^{}]\mbox{, and}
\end{equation}
\begin{equation}
  \label{eq:mu}
  \mu_{\ell ,\nu}^{} \equiv f_\nu \tau_{\ell 0, \nu}
  \left(2 + \frac{\dd \ln \tau_{\ell 0, \nu}}{\dd \ln \ell}\right) = 
  f_\nu \left[2 \tau_{\ell 0, \nu} 
    + \ell\left(\tau_{\ell 0, \nu} - \tau_{\ell-1\, 0, \nu}\right) \right].
\end{equation}
As motivated in Appendix~\ref{sec:app:SAF}, the logarithmic derivative of $\tau_{\ell 0}$ with respect to the multipole order 
$\ell$ is a shorthand notation of the differential quotient which is only valid for $\ell\gg 1$. The quantities $\alpha$ and 
$\beta_\nu$ are given by the components 
\begin{equation}
  \label{eq:components}
  \alpha = (\hat{\bld{A}}^{-1})_{00}, \qquad
  \beta_\nu = (\hat{\bld{A}}^{-1})_{\nu0},
\end{equation}
where $\hat{\bld{A}}$ is the $(1+N) \times (1+N)$ matrix with elements
\begin{eqnarray}
  \label{eq:Amatrix}
  \lefteqn{
  A_{00}^{}\equiv \sum_{\ell=0}^\infty \bld{F}_{\ell}^\rmn{T}
  \hat{\bld{C}}_{\ell}^{-1} \bld{F}_{\ell}^{}, \qquad
  A_{0\nu}^{}\equiv \sum_{\ell=0}^\infty \mu_{\ell ,\nu}^{}
  \left(\bld{F}_{\ell }^\rmn{T}\hat{\bld{C}}_{\ell}^{-1}\right)_\nu}\\
  \lefteqn{
  A_{\nu 0}^{}\equiv \sum_{\ell=0}^\infty \mu_{\ell,\nu}^{}
  \left(\hat{\bld{C}}_{\ell}^{-1}\bld{F}_{\ell }^{}\right)_\nu, \qquad
  A_{\nu \nu'}^{}\equiv \sum_{\ell=0}^\infty 
    \mu_{\ell,\nu}^{}\,\mu_{\ell,\nu'}^{}
    \left(\hat{\bld{C}}_{\ell}^{-1}\right)_{\nu\nu'}^{}}.
\end{eqnarray}
In these equations, no summation over the indices is implied.

\subsection{Detection level and gain}\label{filter_gain_reliability}
As described by \citet{2001ApJ...552..484S}, the concept of constructing an optimised filter function for source detection 
aims at maximising the signal-to-noise ratio $D_u$,
\begin{equation}
  \label{eq:detlevel}
  D_u\equiv\frac{\left\bra u(R_1,\ldots,R_N;\bld{0})\right\ket}{\sigma_u(R_1,\ldots,R_N)}
  = A\cdot\frac{\sum_{\ell=0}^\infty \bld{F}_{\ell }\bpsi_{\ell }}
  {\sqrt{\sum_{\ell=0}^\infty \bpsi_{\ell }^T \hat{\bld{C}}_{\ell}^{} \bpsi_{\ell }^{}}}.
\end{equation}

Computing the dispersion of the unfiltered field on the sphere yields the signal-to-noise ratio $D_s$ of a signal on the 
fluctuating background:
\begin{equation}
  \label{eq:disp}
  \sigma_s^2 = \sum_{\nu_1,\nu_2}\sum_{\ell=0}^\infty C_{\ell,\,\nu_1\nu_2} 
  \quad \Rightarrow \quad D_s = \frac{A}{\sigma_s}.
\end{equation}
These considerations allow introducing the {\em gain} for comparing the signal-to-noise ratios of a peak before and after 
convolution with a filter function:
\begin{equation}
  \label{eq:gain}
  g \equiv \frac{D_u}{D_s} = \frac{\sigma_s}{\sigma_u(R_1,\ldots,R_N)}.
\end{equation}

If the noise suppression is successful, the gain $g$ will assume values larger than one. If the filters are constructed 
efficiently, they are able to reduce the dispersion ($\sigma_u(R_1,\ldots,R_N)<\sigma_s$) while simultaneously retaining the 
expectation value of the field (\ref{eq:umean}). Due to the additional third constraint, the scale-adaptive filter is expected 
to achieve smaller gains compared to the matched filter.

\subsection{Numerical derivation of filter kernels}\label{filter_numerics}
For the derivation of suitable filter kernels the source profiles are assumed to be generalised King-profiles as described by 
eqn.~(\ref{eqn_profile_beam_source}) convolved with the respective \planck-beam superimposed on fluctuating background given by 
template $\bra S_{\ell m}\ket_\nu$-coefficients. The inversion of the matrix $\hat{\bld{C}}_{\ell}$ (c.f. 
eqns.~(\ref{eqn_cross_power}) and (\ref{eqn_cross_power_def})) can be performed using either Gauss-Jordan elimination or LU 
decomposition, which both were found to yield reliable results. In the derivation of the scale-adaptive filters, however, it is 
numerically advantageous to artificially exclude the lower multipoles $\ell\leq1$ from the calculation. Due to the 
sub-millimetric emission of the Milky Way, the lower multipoles are very large. Consequently, the corresponding $\psi_{\ell 
m}$-coefficients, $\ell\leq1$, have been set to zero, which is not a serious intervention since the filters are designed to 
amplify structures at angular scales well below a degree. For consistency, the multipoles below the quadrupole have been 
artificially removed in the derivation of the matched filters as well.

In contrast to the \planck-simulation pipeline all numerical calculations presented here are carried out in terms of fluxes 
measured in Jy and not in antenna temperatures for the following reason: Cross-power spectra $C_{\ell, \nu_1\nu_2}$ 
given in terms of antenna temperatures are proportional to $(\nu_1\cdot\nu_2)^{-2}$ which results in a suppression of the 
highest frequency channels by a factor of almost $10^5$ compared to the lowest frequency channels.

Furthermore, by working with fluxes instead of antenna temperatures, the filters for extracting the SZ-signal show frequency 
dependences which can be understood intuitively. The frequency dependence is described by eqns.~\ref{eq:S_thSZ} and 
\ref{eq:S_kinSZ}. The normalisation $\mathcal{Y}$ has been chosen to be $1~\mathrm{arcmin}^2$, which corresponds to the weakest 
signals \planck~ will be able to detect. Because of the smallness of the source profiles to be detected, the calculations were 
carried out to multipole orders of $\ell_\mathrm{max}=4096$, which ensures that the beams as well as the source profiles are 
well described. In the plots in Sect.~\ref{filter_shape_discussion}, the filters depicted are smoothed with a moving average 
window comprising eleven bins for better visualisation.

\subsection{Discussion of filter kernels}\label{filter_shape_discussion}

\subsubsection{Matched filter}

\begin{figure*}
\begin{tabular}{cc}
\resizebox{0.48\linewidth}{!}{\includegraphics{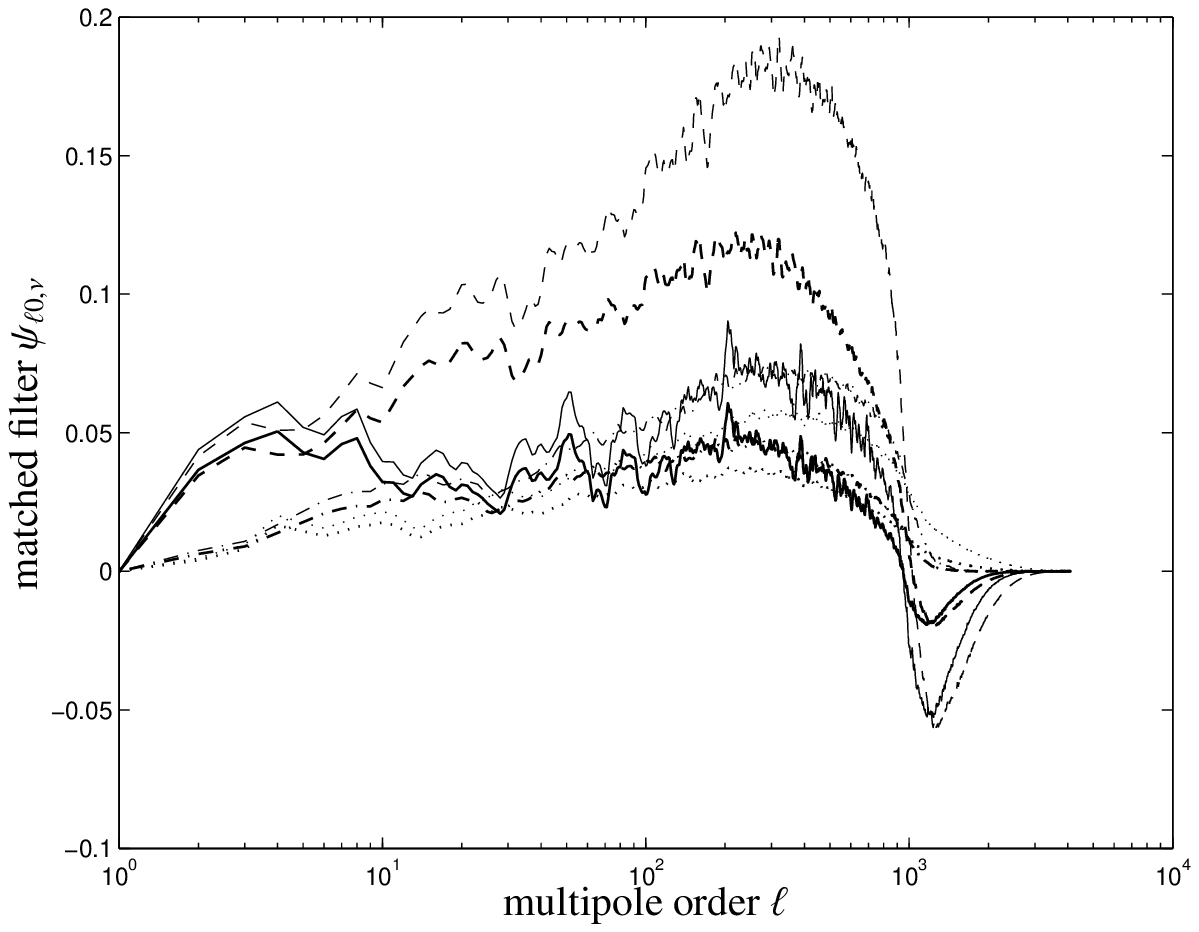}} &
\resizebox{0.48\linewidth}{!}{\includegraphics{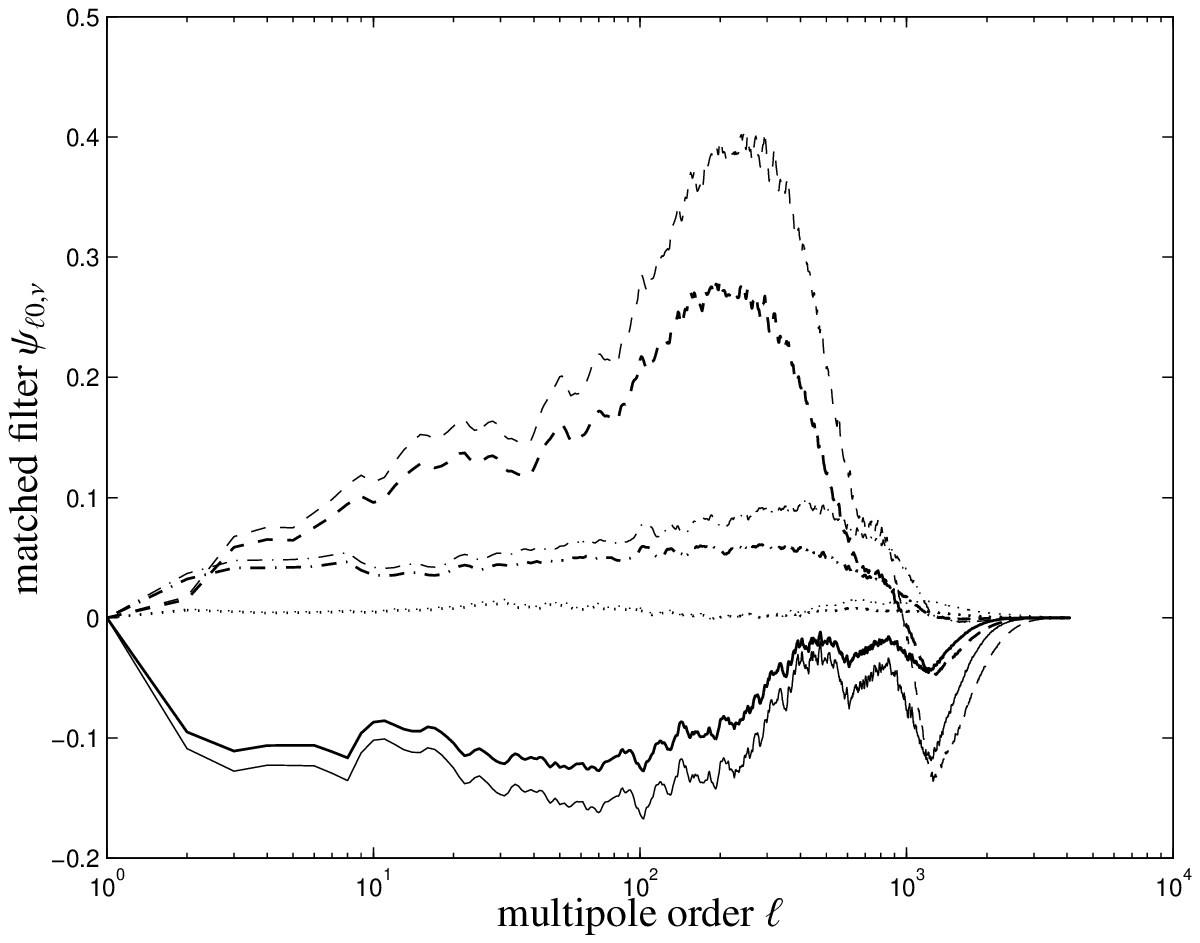}} \\
\resizebox{0.48\linewidth}{!}{\includegraphics{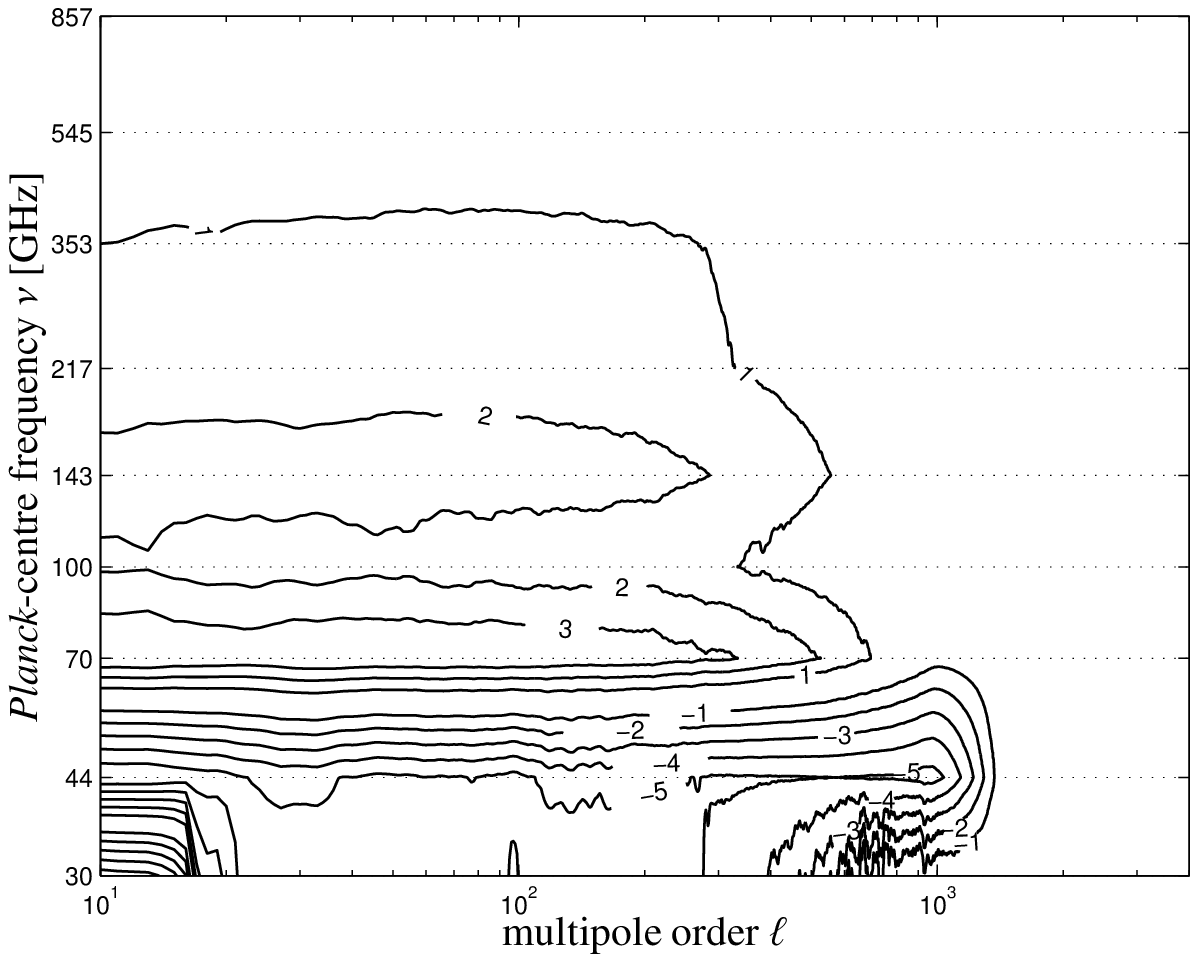}} &
\resizebox{0.48\linewidth}{!}{\includegraphics{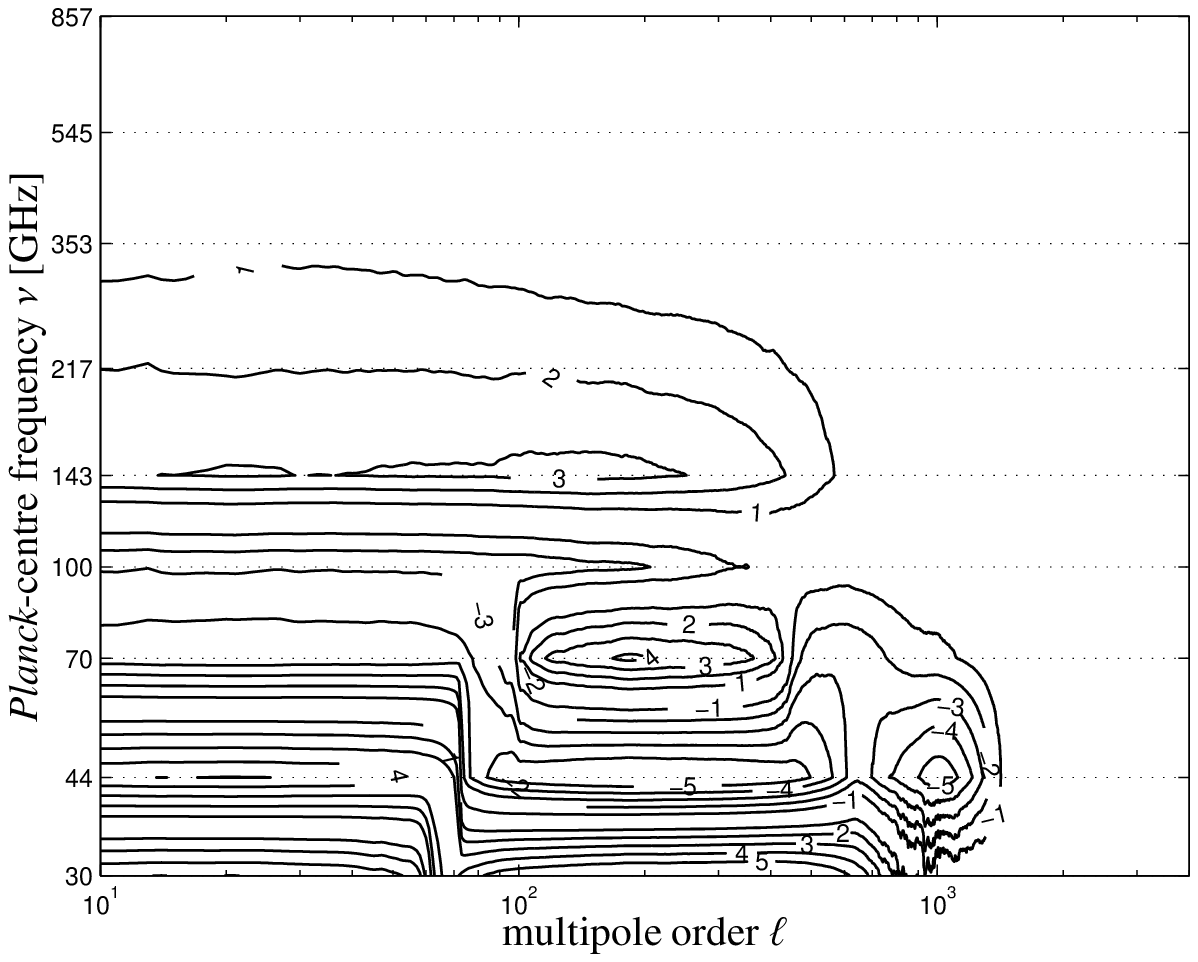}}
\end{tabular}
\caption{{\em Upper panel:} Spherical harmonics coefficients $\psi_{\ell 0,\nu}$ as derived with the matched filter 
algorithm, for $\nu=100$~GHz (solid line), $\nu=143$~GHz (dashed line), $\nu=217$~GHz (dash-dotted line) and $\nu=353$~GHz 
(dotted line). The left and right columns compare filter kernels derived for a data set containing only the CMB, the SZ-effects 
and instrumental noise (left column) and for a data set comprising all Galactic foregrounds in addition (right column). The 
filter kernels are optimised for the detection of generalised King-profiles with core radii $\theta_c = 3\farcm0$ (thin lines), 
$\theta_c=5\farcm0$ (thick lines) and asymptotic slope $\lambda=1.0$.\newline
{\em Lower panel: } Contour plot of the spherical harmonics expansion coefficients $\psi_{\ell 0,\nu}$ derived with the 
matched filter algorithm as a function of both the multipole moment order $\ell$ and \plancks observing frequency $\nu$. The 
filter kernels have been derived for a optimised detection of a generalised King-profile with 
$(\theta_c,\lambda)=(15\farcm0,1.0)$. The contours are linearly spaced in $\mathrm{arsinh}(10^2\cdot\psi_{\ell 0,\nu})$. 
}
\label{figure_filter_kernel_matched}
\end{figure*}

The spherical harmonics expansion coefficients $\psi_{\ell 0,\nu}$ following from the matched filter algorithm are depicted in 
Fig.~\ref{figure_filter_kernel_matched} for four frequencies most relevant to SZ-observation, namely for $\nu=100$~GHz, 
$\nu=143$~GHz, $\nu=217$~GHz and $\nu=353$~GHz. As background noise components the clean {\tt COS} data set (left column) and 
the exhaustive {\tt GAL} data set (right column) are contrasted. The filter kernels have been derived for optimised detection 
of sources described by a generalised King-profile with angular core radii $\theta_c=3\farcm0$ and $\theta_c=5\farcm0$ and 
asymptotic slope $\lambda=1.0$.

The principle how the matched filter extracts the SZ-signal from the maps is explained by 
Fig.~\ref{figure_filter_kernel_matched}: The SZ-profiles the filter has been optimised are small structures at 
angular scales corresponding to multipole moments of $\ell\simeq 10^3$. In channels below $\nu=217$~GHz, the clusters are 
observed in absorption and the fluxes are decreased. For that reason, the filters have negative amplitudes at these angular 
scales for these specific frequencies. At larger scales, the fluctuations are suppressed by linear combination of the various 
channels, while the filtering functions show very similar shapes. Optimising the filters for detection of core radii of 
$5\farcm0$ instead of $3\farcm0$ result in a shift of the negative peak at $\ell\simeq10^3$ to smaller multipole orders. 
Instrumental noise which is important at even higher multipoles is suppressed by the filter's exponential decline at high 
$\ell$ above $\ell\gsim 2000$. The unwanted CMB fluctuations and all Galactic contributions at scales larger than the cluster 
scale are suppressed by weightings with varying sign so that the foregrounds are subtracted at the stage of forming linear 
combinations of the $\bra S_{\ell m}\ket_\nu$-coefficients.

Furthermore, the contours of the matched filter kernels are given in Fig.~\ref{figure_filter_kernel_matched} as functions of 
both inverse angular scale $\ell$ and observing frequency $\nu$ for differing noise contributions. The figures compare filters 
derived for differing background noise compositions. The filters shown serve for the optimised detection of generalised 
King-profiles with core radius $\theta_c=15\farcm0$ and asymptotic slope $\lambda=1.0$. These (rather large) values have been 
chosen for visualisation purposes. For clarity, the contour denoting zero values has been omitted due to noisy data. In these 
figures it is apparent how the filters combine the frequency information in order to achieve a suppression of the unwanted 
foregrounds: At multipole moments of a few hundred, the filters exhibit changes in sign, such that the measurements at low 
frequencies are subtracted from the measurements at high frequencies in the linear combination of the filtered maps.

\begin{figure}
\resizebox{\hsize}{!}{\includegraphics{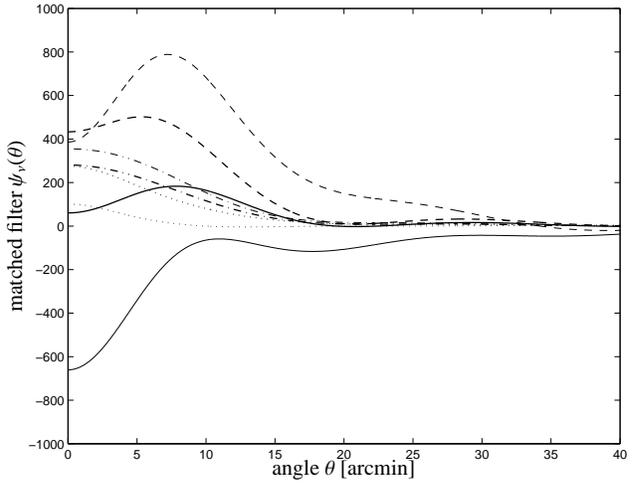}}
\caption{Matched filter kernels $\psi_\nu(\theta)$ in real space at SZ-frequencies, i.e. for $\nu=100$~GHz (solid line), 
$\nu=143$~GHz (dashed line), $\nu=217$~GHz (dotted line) and $\nu=353$~GHz (dash-dotted line), for a data set including the 
CMB, Galactic foregrounds and instrumental noise (thin lines) and for a data set containing all Galactic components in addition 
to the CMB and instrumental noise (thick lines). The filter kernel is optimised for the detection of a generalised King-profile 
with core radius $\theta_c = 5\farcm0$ and asymptotic slope $\lambda=1.0$.}
\label{figure_filter_kernel_matched_real}
\end{figure}

Fig.~\ref{figure_filter_kernel_matched_real} illustrates the filter kernels $\psi_\nu(\theta)$ in real space for the same 
selection of frequencies and background noise components as given above. The filter kernels $\psi_\nu(\theta)$ have been 
synthesised from the $\psi_{\ell 0,\nu}$-coefficients using the {\tt alm2grid}-utility of the \planck-simulation package. Here, 
the parameters of the King-profile to be detected are $(\theta_c,\lambda)=(5\farcm0,1.0)$. The filter kernels are similar in 
shape to Mexican-hat wavelets, but show more than one oscillation. Their action on the sky maps is to apply high-pass 
filtering, such that all long-wavelength modes are eliminated. At the cluster scale, they implement a linear combination of the 
sky maps that aims at amplifying the SZ-signal: The kernels derived for both the $\nu=100$~GHz- and $\nu=143$~GHz-channel 
exhibit a central depression which is used to convert the SZ-signal to positive amplitudes. The other two channels resemble 
simple Gaussian kernels which smooth the maps to a common effective angular resolution. At frequencies of $\nu=217$~GHz and 
$\nu=353$~GHz the most important emission feature is Galactic Dust, which is suppressed by the filter's small amplitudes. In 
this way, the weak SZ-signal is dissected.

\begin{figure}
\resizebox{\hsize}{!}{\includegraphics{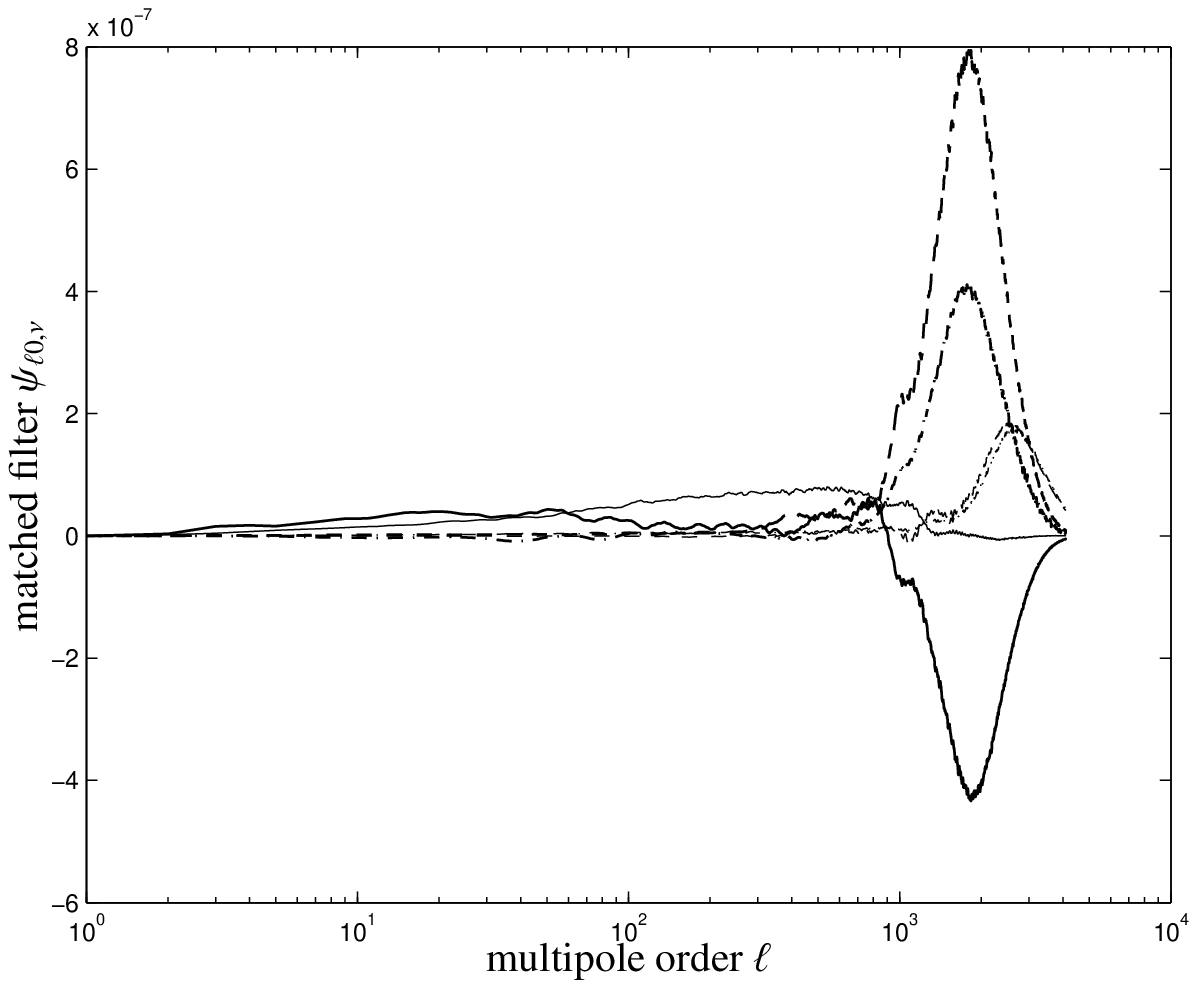}}
\caption{Comparison of filter kernel $\psi_{\ell 0,\nu}$-coefficients derived for differing spectral dependences of the signal: 
thermal SZ-effect (solid line), kinetic SZ-effect (dashed line) and a Planckian emitter with surface temperature of 
$T_\mathrm{surface} = 150~\mathrm{K}$ (dash-dotted line). All sources are assumed to be point-like, i.e. they appear to have 
the shape of the \planck-beam. The curves are given for observing frequencies of $\nu=143$~GHz (thin line) and $\nu=217$~GHz 
(thick line) and have been derived with the matched filter algorithm. The noise is a composite of CMB fluctuations, 
Galactic and ecliptic foregrounds and instrumental noise.}
\label{figure_filter_kernel_matched_spec}
\end{figure}

In Fig.~\ref{figure_filter_kernel_matched_spec}, filter kernels derived with both algorithms for point sources (i.e. with 
beam profiles of the respective \planck-channels) are compared, that have been optimised for the detection of varying spectral 
behaviour of the signal, in this case the thermal SZ-effect, the kinetic SZ-effect and a Planckian thermal emitter with a 
surface temperature $T_\mathrm{surface}$ of 150~K, such as an asteroid or planet. The filter kernels depicted correspond to  
observing frequencies of $\nu=143$~GHz and $\nu=217$~GHz. The filters clearly reflect the spectral behaviour of the emission 
laws of the sources one aims at detecting: While the filter kernels designed for detecting thermal SZ-clusters reflect the 
peculiar change in sign in the SZ-effect's frequency dependence, the other two curves show the behaviour to be expected for a 
Planckian emitter and the kinetic SZ-effect, respectively. Again, the better angular resolution of the $\nu=217$~GHz-channel is 
apparent by the shifting of the curves to higher multipole order $\ell$.

\subsubsection{Scale-adaptive filter}

\begin{figure*}
\begin{tabular}{cc}
\resizebox{0.48\linewidth}{!}{\includegraphics{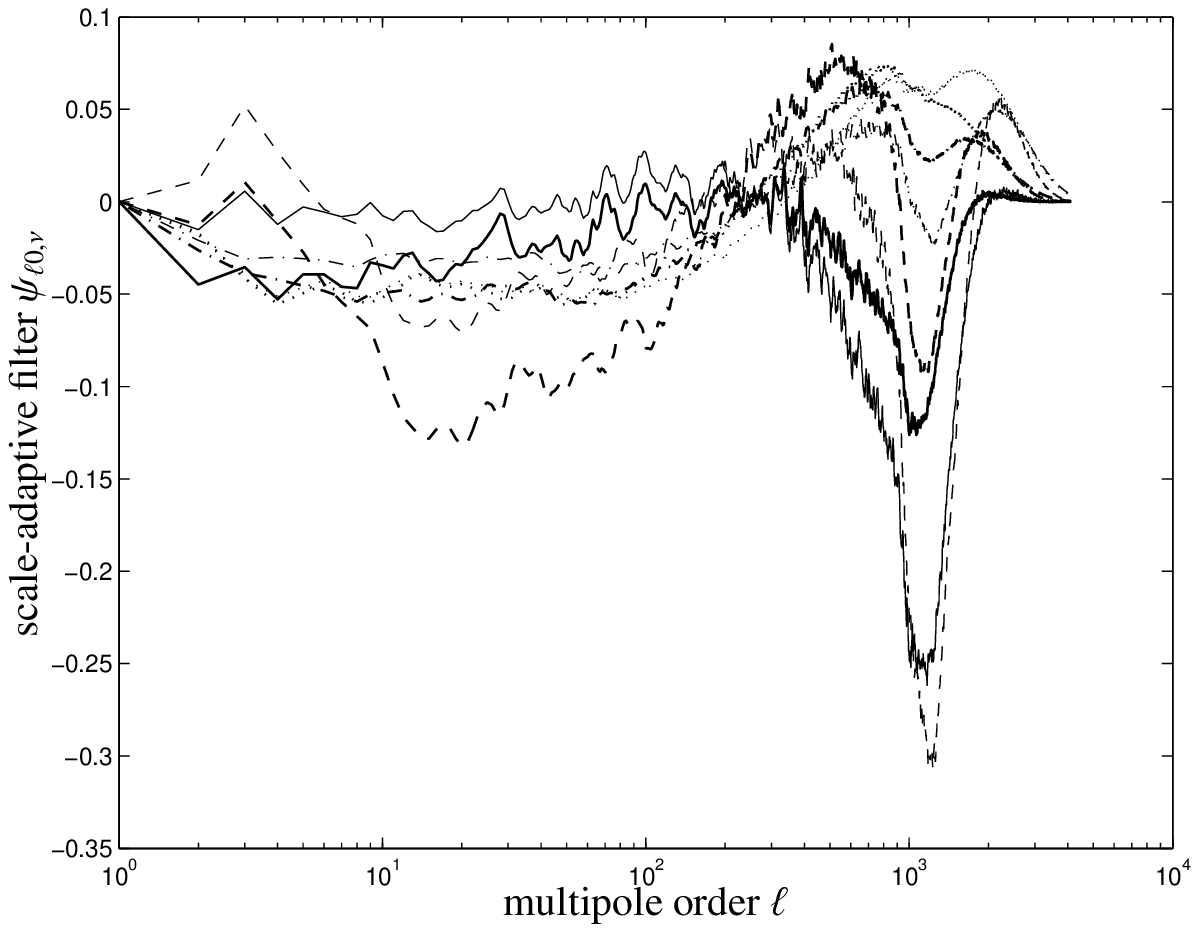}} &
\resizebox{0.48\linewidth}{!}{\includegraphics{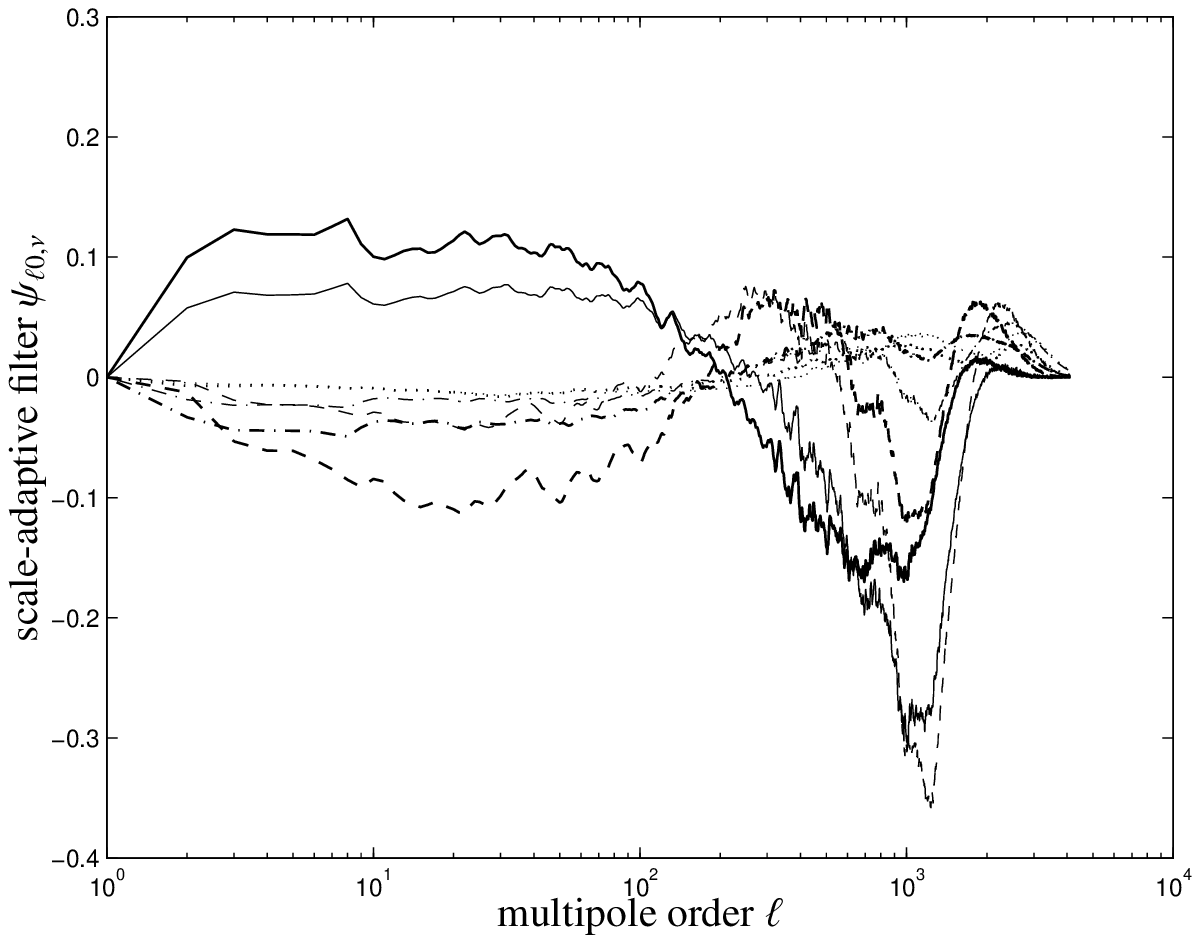}} \\
\resizebox{0.48\linewidth}{!}{\includegraphics{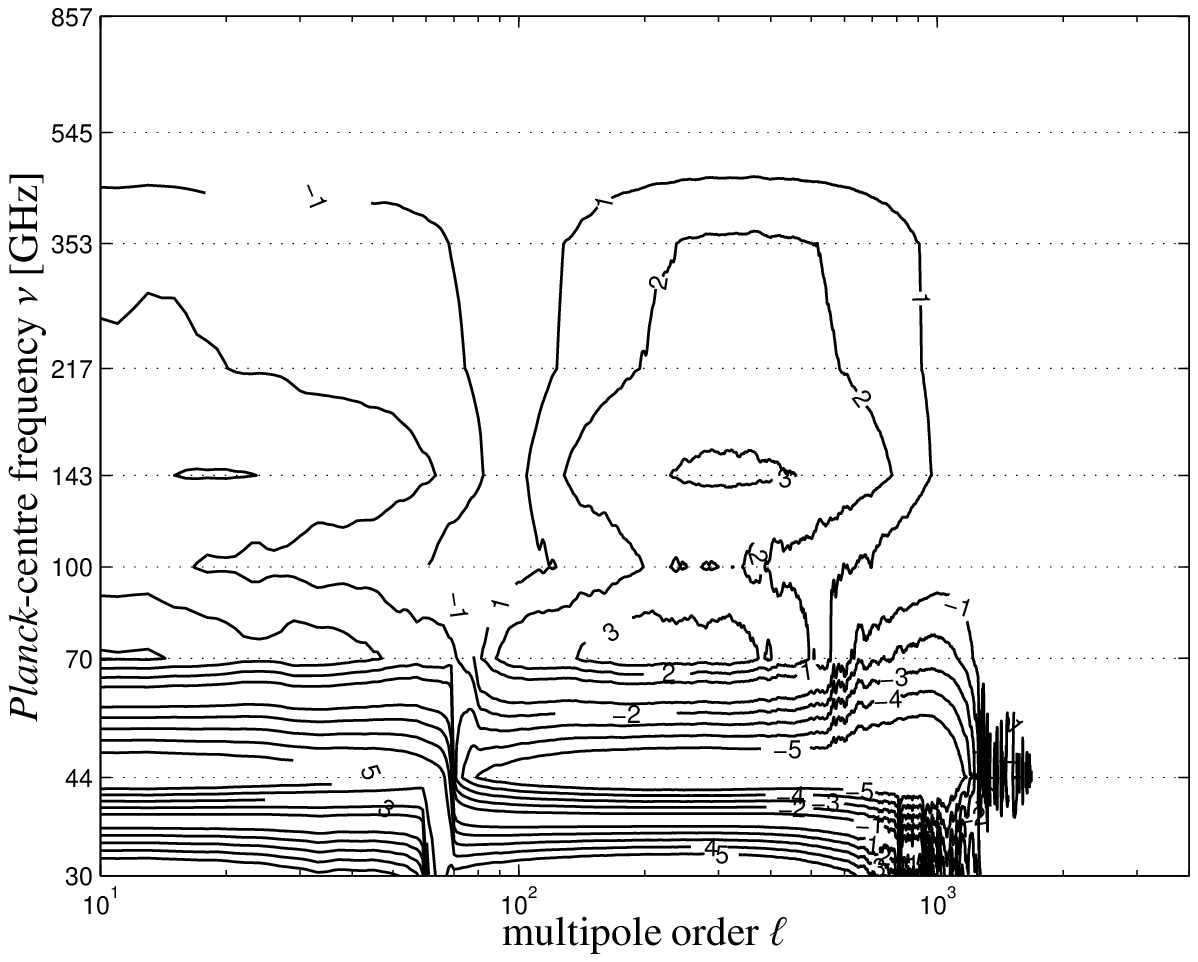}} &
\resizebox{0.48\linewidth}{!}{\includegraphics{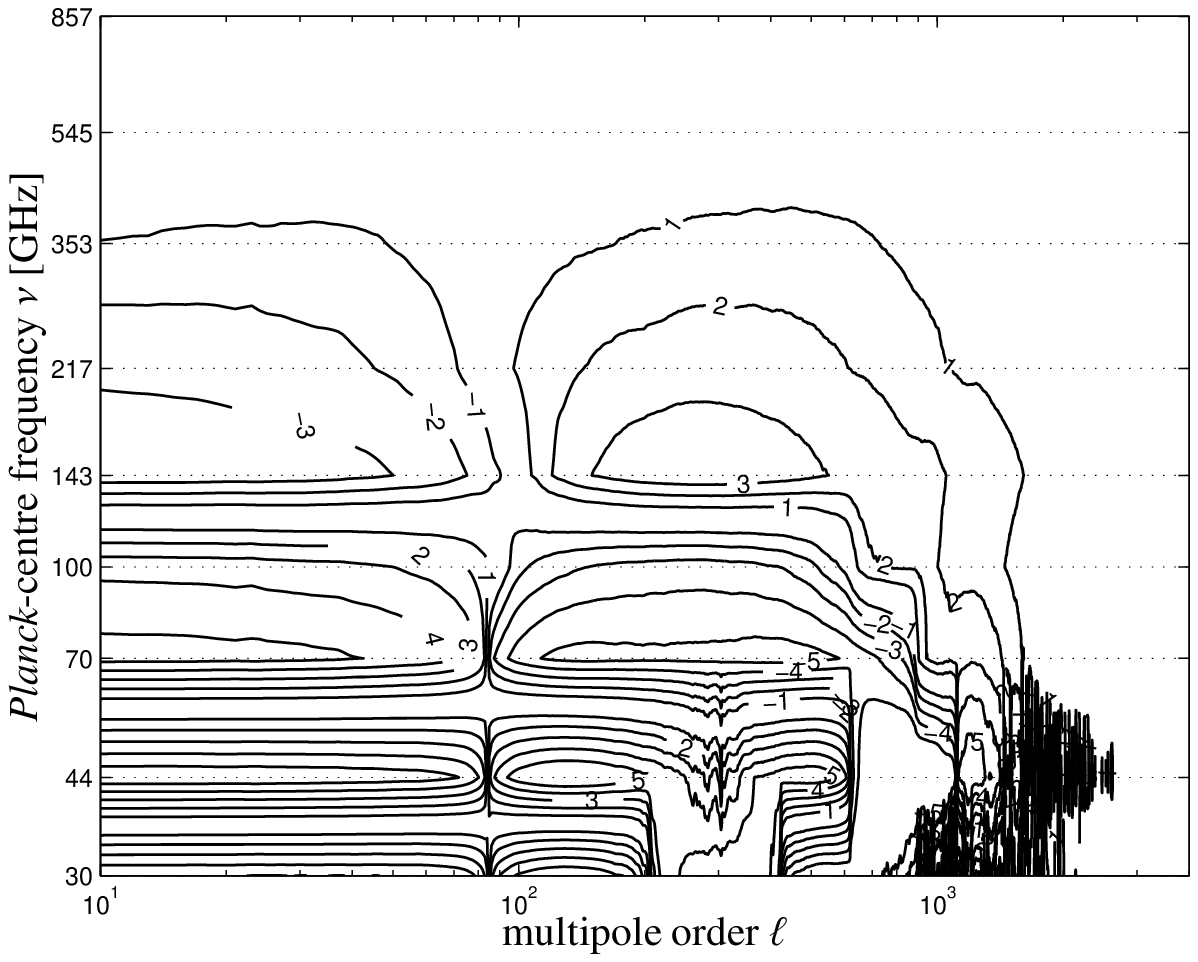}}
\end{tabular}
\caption{{\em Upper panel:} Spherical harmonics coefficients $\psi_{\ell 0,\nu}$ as derived with the scale-adaptive filter 
algorithm, for $\nu=100$~GHz (solid line), $\nu=143$~GHz (dashed line), $\nu=217$~GHz (dash-dotted line) and $\nu=353$~GHz 
(dotted line). The filter kernel is optimised for the detection of generalised King-profiles with core radii $\theta_c = 
3\farcm0$ (thin lines) and $\theta_c=5\farcm0$ (thick lines) and asymptotic slope $\lambda=1.0$.\newline
{\em Lower panel:} Contour plots of the spherical harmonics expansion coefficients $\psi_{\ell 0,\nu}$ derived with the 
scale-adaptive filter algorithm as a function of both the multipole moment order $\ell$ and \plancks observing frequency $\nu$ 
are shown. The filter kernels have been derived for a optimised detection of a generalised King-profile with 
$(\theta_c,\lambda)=(15\farcm0,1.0)$. The contours are linearly spaced in $\mathrm{arsinh}(10^2\cdot\psi_{\ell 0,\nu})$. Again, 
the columns compare the influence of the fluctuating background, comprising cosmological contributions (CMB and both 
SZ-effects) and instrumental noise (left column) to the data set that contains all Galactic foregrounds in addition.}
\label{figure_filter_kernel_adaptive}
\end{figure*}

The spherical harmonics expansion coefficients $\psi_{\ell 0,\nu}$ following from the scale-adaptive filter algorithm for the 
frequencies $\nu=100$~GHz, $\nu=143$~GHz, $\nu=217$~GHz and $\nu=353$~GHz are given in the upper panel of
Fig.~\ref{figure_filter_kernel_adaptive}. The left and right columns compare the filter kernels for differing 
noise components. Their functional shape has a number of important features in common with the matched filters: They suppress 
the uncorrelated pixel noise, which is dominant at high $\ell$ by their exponential decline at $\ell\gsim2000$. Furthermore, 
the filters amplify the SZ-signal, which is negative at frequencies below $\nu=217$~GHz, by assuming large negative values and 
hence converting the SZ-signal to yield positive amplitudes. Additionally, the filters show a distinct secondary peak at 
$\ell\simeq2000$ which causes the kernels to be more compact after transformation to real space and enables the size 
measurement. A more general observation is that the scale-adaptive filter kernel shapes are more complex and  noisier 
in comparison to the matched filter, especially at high $\ell$. 

The scale-adaptive filter makes even stronger use of the spectral information than the matched filter. Especially the contour 
plots in Fig.~\ref{figure_filter_kernel_adaptive} show that the scale-adaptive filter exhibits alternating signs when varying 
the observing frequency $\nu$ while keeping the angular scale $\ell$ fixed. In this way, the noise contributions are isolated 
in angular scale and subsequently suppressed by linear combination of the maps. Furthermore, one notices a change in sign at 
multipole order $\ell\simeq 200$ which is common to the frequencies $\nu=100\ldots353$~GHz, at which the CMB signal is 
strongest. Aiming at reducing the variance of the filtered maps, the scale-adaptive filter is suppressing the $\bra S_{\ell 
m}\ket_\nu$-coefficients by assuming small values.

\begin{figure}
\resizebox{\hsize}{!}{\includegraphics{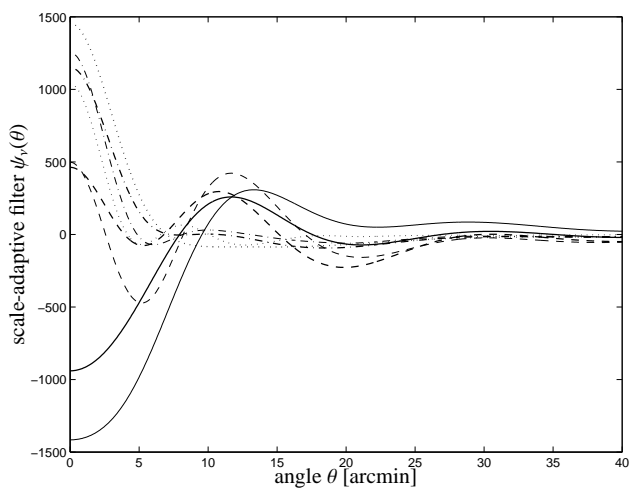}}
\caption{Scale-adaptive filter kernels $\psi_\nu(\theta)$ in real space, for $\nu=100$~GHz (solid line), $\nu=143$~GHz (dashed 
line), $\nu=217$~GHz (dotted line) and $\nu=353$~GHz (dash-dotted line), for a data set incorporating the CMB, Galactic 
foregrounds and instrumental noise. The filter kernel is optimised for the detection of a generalised King-profile with 
parameters $(\theta_c,\lambda) = (5\farcm0,1.0)$.}
\label{figure_filter_kernel_adaptive_real}
\end{figure}

Fig.~\ref{figure_filter_kernel_adaptive_real} gives the filter kernels $\psi_\nu(\theta)$ in real space for selected 
frequencies and background noise components. The scale-adaptive filters work similarly as the matched filters like Mexican-hat 
wavelets and subject the sky maps to high pass filtering. 

\begin{figure}
\resizebox{\hsize}{!}{\includegraphics{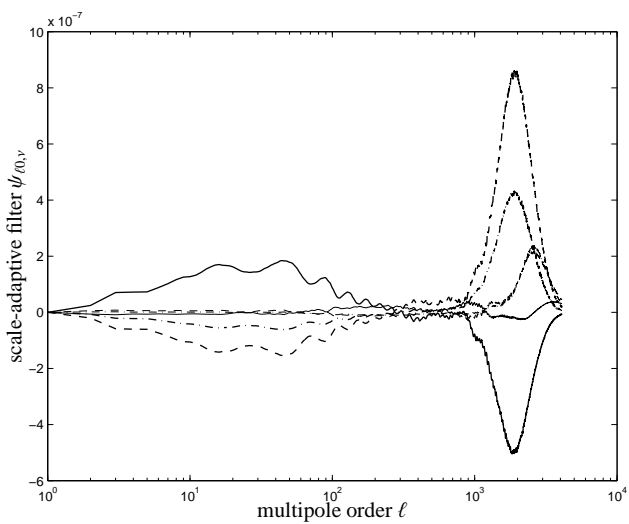}}
\caption{Comparison of filter kernel $\psi_{\ell 0,\nu}$-coefficients derived for differing spectral dependences of the signal: 
thermal SZ-effect (solid line), kinetic SZ-effect (dashed line) and a Planckian emitter with surface temperature of 
$T_\mathrm{surface} = 150~\mathrm{K}$ (dash-dotted line). All sources are assumed to be point-like, i.e. they appear to have 
the shape as the \planck-beam. The curves are given for observing frequencies of $\nu=143$~GHz (thin line) and $\nu=217$~GHz 
(thick line) and have been derived with the scale-adaptive filter algorithm. The noise is a composite of CMB fluctuations, 
Galactic and ecliptic foregrounds and instrumental noise.}
\label{figure_filter_kernel_adaptive_spec}
\end{figure}

In Fig.~\ref{figure_filter_kernel_adaptive_spec}, filter kernels derived with both algorithms for point sources (i.e. with 
beam profiles of the respective \planck-channels) are compared, that have been optimised for the detection of varying spectral 
behaviour of the signal, in this case the thermal SZ-effect, the kinetic SZ-effect and a Planckian thermal emitter with a 
surface temperature $T_\mathrm{surface}$ of 150~K, such as an asteroid or planet. The filter kernels depicted correspond to  
observing frequencies of $\nu=143$~GHz and $\nu=217$~GHz. As in the case of the matched filter, the frequency dependence of the 
signal is reflected by the sign of the filter kernel at the anticipated angular scale of the profile to be detected.

\subsection{Filter renormalisation and synthesis of likelihood maps}\label{filter_renormalise}
Once the filter kernels are derived, the filtered fields $u_{\nu}(R_\nu,\bmath{\beta})$ can be synthesised from the $u_{\ell 
m,\nu}$-coefficients (defined in eqn.~(\ref{eqn_k_convolve})) and the resulting maps can be added in order to yield the 
co-added, filtered field $u(R_1,\ldots,R_N,\bmath{\beta})$ (see eqn.~(\ref{eqn_k_add})), which can be normalised by the level 
of fluctuation $\sigma_u$ (given by eqn.~(\ref{eqn_def_ccspec})) to yield the likelihood map $D(\bmath{\theta})$. It is 
favourable to divide the filter kernels by the variance $\sigma_u$ and to apply a renormalisation:
\begin{equation}
\psi_{\ell 0,\nu}\longrightarrow\psi^\prime_{\ell 0,\nu} = 
\frac{\psi_{\ell 0,\nu}}{\sqrt{\sum_\ell \bmath{\psi}_{\ell}^T\hat{\bld{C}}_{\ell}\bmath{\psi}_{\ell}}}\mbox{.}
\end{equation}
In this case, the filter kernels are invariant under changes in profile normalisation. With these kernels, the filtered flux 
maps can be synthesised from the set of $\bra S_{\ell m}\ket_\nu$-coefficients and the resulting maps can be co-added to yield 
the final normalised likelihood map $D(\bmath{\beta})$. It is computationally advantageous, however, to interchange the last 
two steps,
\begin{eqnarray}
D_u(\bmath{\beta}) & = & \frac{u(\bmath{\beta})}{\sigma_u} = \frac{1}{\sigma_u}\sum_\nu u_\nu(\bmath{\beta})\\
 & = & 
\sum_\nu\sum_{\ell=0}^{\infty}\sum_{m=-\ell}^{+\ell} 
\sqrt{\frac{4\pi}{2\ell+1}}\bra S_{\ell m}\ket_\nu\frac{\psi_{\ell 0,\nu}}{\sqrt{\sum_\ell 
\bmath{\psi}^T_\ell\hat{\bld{C}}_\ell\bmath{\psi}_\ell}} Y_\ell^m(\bmath{\beta})\\
& = & \sum_{\ell=0}^{\infty}\sum_{m=-\ell}^{+\ell}\underbrace{\sqrt{\frac{4\pi}{2\ell+1}}\left[\sum_\nu \bra S_{\ell m}\ket_\nu 
\cdot\psi^{\prime}_{\ell 0,\nu}\right]}_{\equiv D_{\ell m}} Y_\ell^m(\bmath{\beta})\mbox{,}
\end{eqnarray}
and to derive the $D_{\ell m}$-coefficients first, such that the synthesis has to be performed only once. Due to the 
restriction to axially symmetric kernels, the convolution can be carried out using the {\tt alm2map}-utility rather than 
{\tt totalconvolve}.

\begin{figure*}
\begin{tabular}{cc}
\resizebox{0.47\linewidth}{!}{\frame{\includegraphics{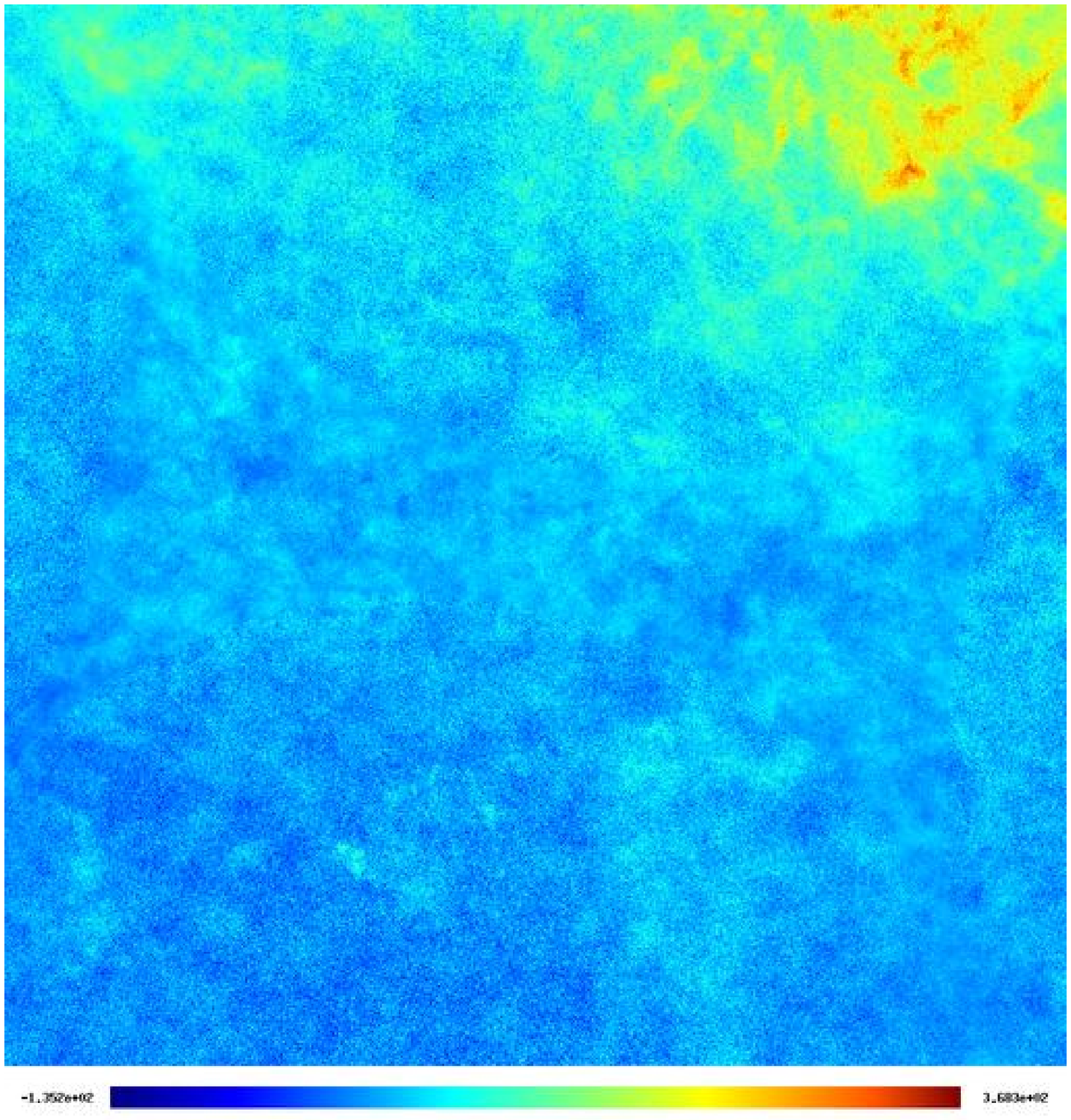}}} & 
\resizebox{0.47\linewidth}{!}{\frame{\includegraphics{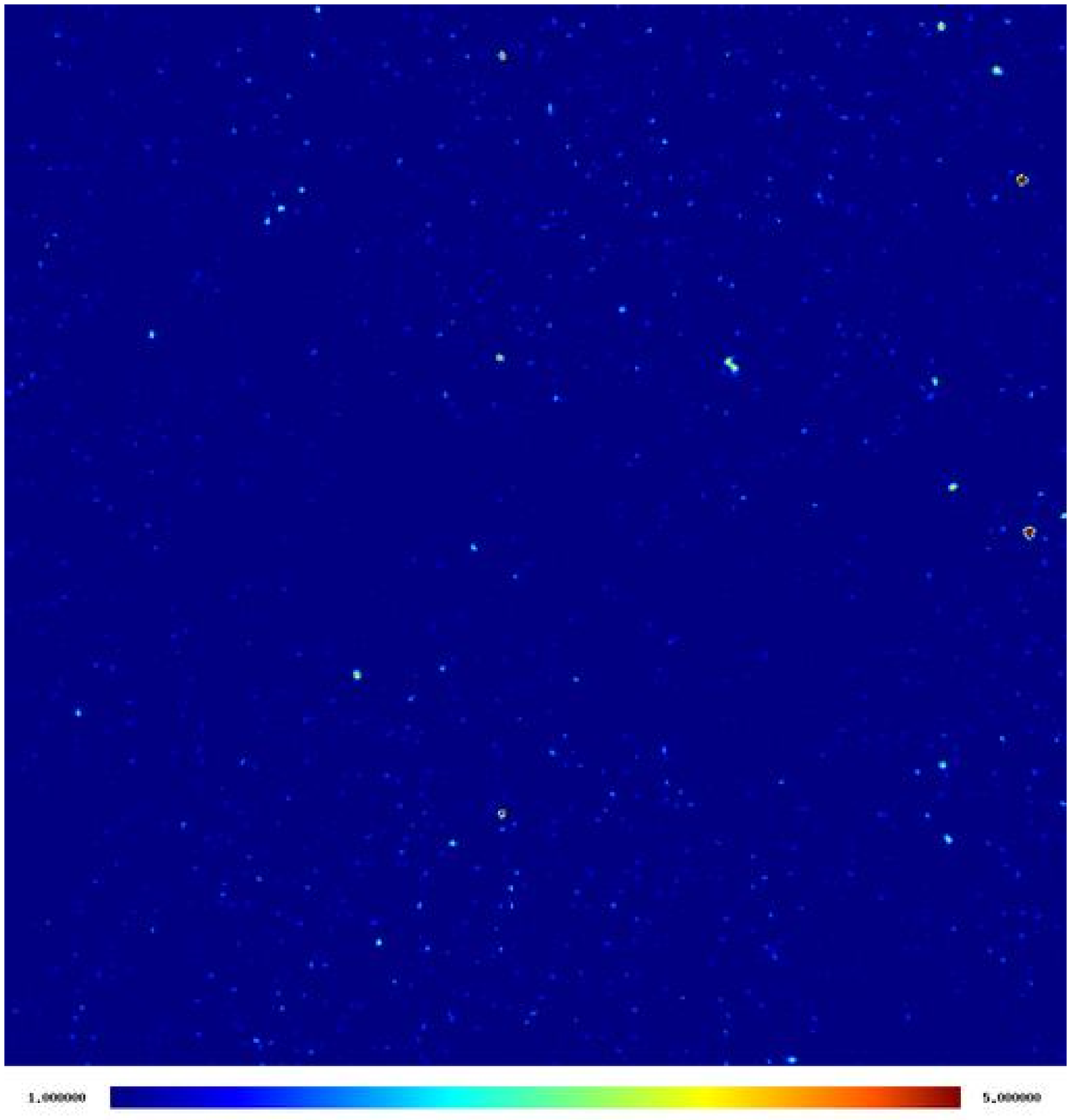}}} \\
\resizebox{0.47\linewidth}{!}{\frame{\includegraphics{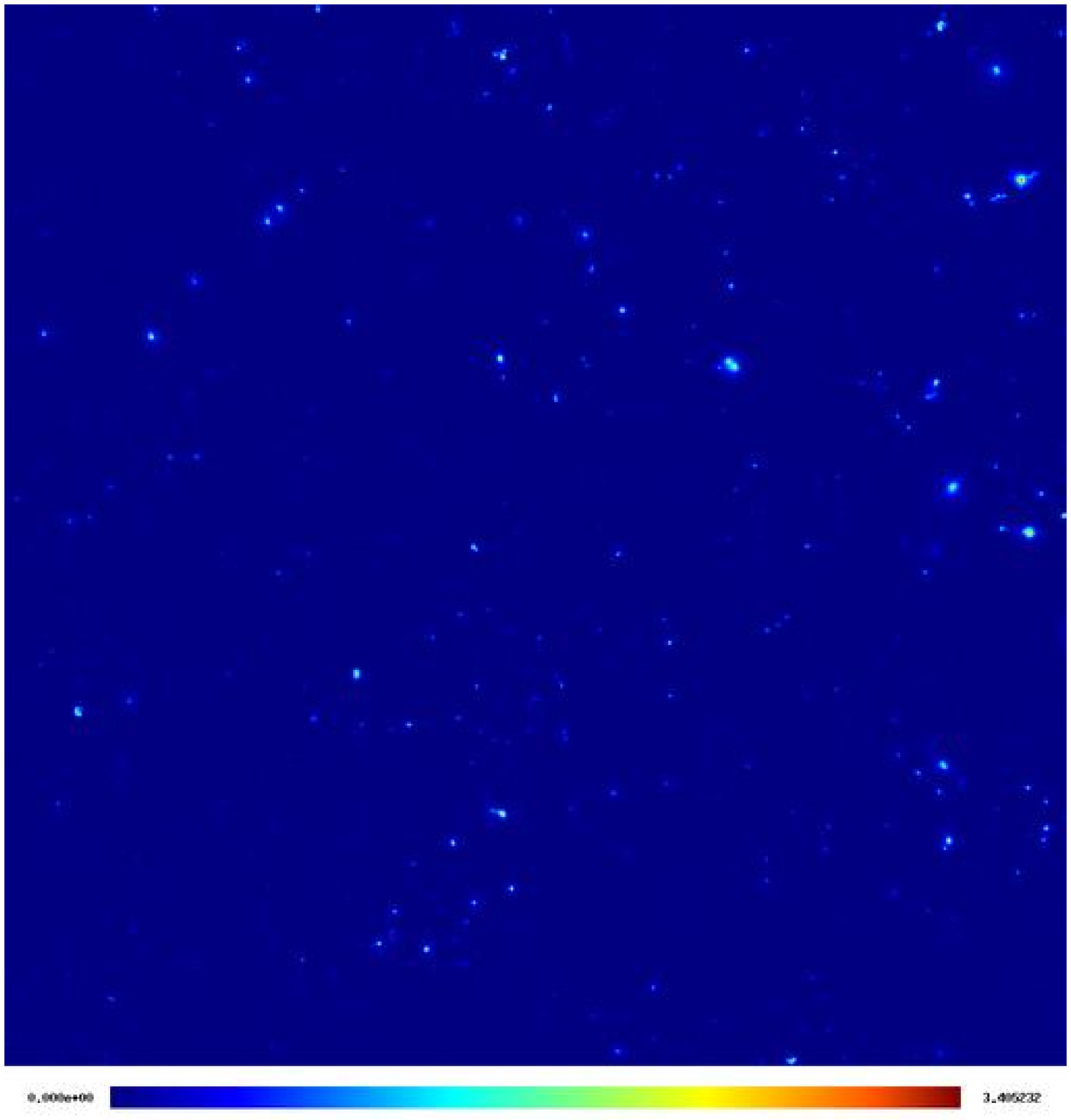}}} & 
\resizebox{0.47\linewidth}{!}{\frame{\includegraphics{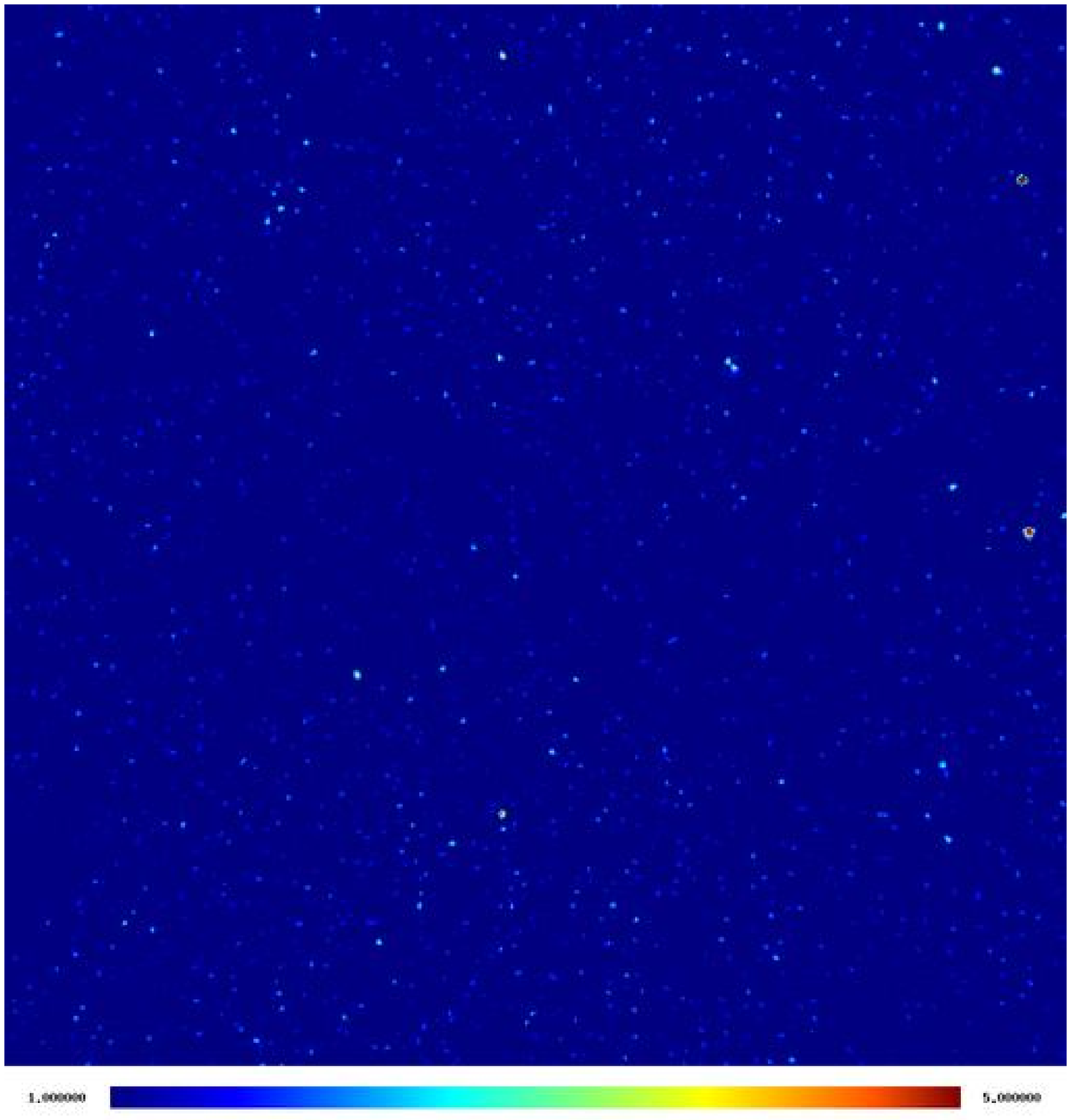}}}
\end{tabular}
\caption{{\em Upper left panel:} A $30\degr\times30\degr$ wide field centered on the ecliptic North pole as synthesised from 
a data set containing the CMB, all Galactic and ecliptic foregrounds and instrumental noise for an observing frequency of 
$\nu=353$~GHz is shown. The amplitudes are proportional to $\mathrm{arsinh}(T_A(\nu=353~\mathrm{GHz})/\mathrm{\umu K})$ and the 
field is smoothed with the corresponding \planck-beam of $\Delta\theta=5\farcm0$ (FWHM). 
{\em Upper right panel:} The same field is shown after reconstruction from the $D_{\ell m}$-coefficients. Here, filters derived 
with the matched filter algorithm optimised for detecting point sources were employed. The amplitudes are given in detection 
significances and the shading scales linearly. {\em Lower right panel:} Again, the same field is shown after synthesis from 
the $D_{\ell m}$-coefficients but in this case, filters derived with the scale-adaptive filter algorithm optimised for 
detecting point sources were used. The amplitudes are stated as detection significances and the shading is linear. In the {\em 
lower left panel}, the corresponding field taken from the original thermal SZ-map is given for comparison. The amplitudes are 
$\propto\mathrm{arsinh(10^4\cdot y)}$.}
\label{figure_filter_likelihood}
\end{figure*}

Fig.~\ref{figure_filter_likelihood} gives a visual impression of the capability of the above described filtering schemes: The 
figure shows a $30\degr\times30\degr$ wide field at the ecliptic North pole at a frequency of $\nu=353$~GHz (at the SZ-maximum) 
as observed by \planck, i.e. the image is smoothed to an angular resolution of $\Delta\theta=5\farcm0$ (FWHM) and contains the 
fluctuating CMB, all Galactic and ecliptic foregrounds as well as pixel noise. Matched and scale-adaptive filter kernels were 
derived for isolating point sources, i.e. for sources that appear to have profiles equal to \plancks beams of the corresponding 
channel. For clarity, only amplitudes exceeding a threshold value of 1.0 are shown. 

For comparison, Fig.~\ref{figure_filter_likelihood} shows the same detail of the input thermal SZ-map as well. It is 
immediately apparent that the observation of SZ-clusters without foreground- and noise suppression is not possible and that one 
has to rely on filtering schemes. As a comparison with Fig.~\ref{figure_filter_likelihood} shows, the filters are clearly able 
to isolate the SZ-clusters and to strongly suppress all spurious noise contributions. The matched filter, however, shows a 
slightly better performance and yields more significant peaks due to better background suppression. There are weak residuals 
present in both maps due to incomplete foreground reduction. These residuals however, have small amplitudes compared to the 
SZ-detections. The highest peaks exhibit detection significances amounting to $10.6\sigma$ in the case of the matched filter 
and $9.1\sigma$ in the case of the scale-adaptive filter. 

\begin{figure*}
\begin{tabular}{cc}
\hspace{-0.5cm}
\resizebox{!}{10cm}{\includegraphics{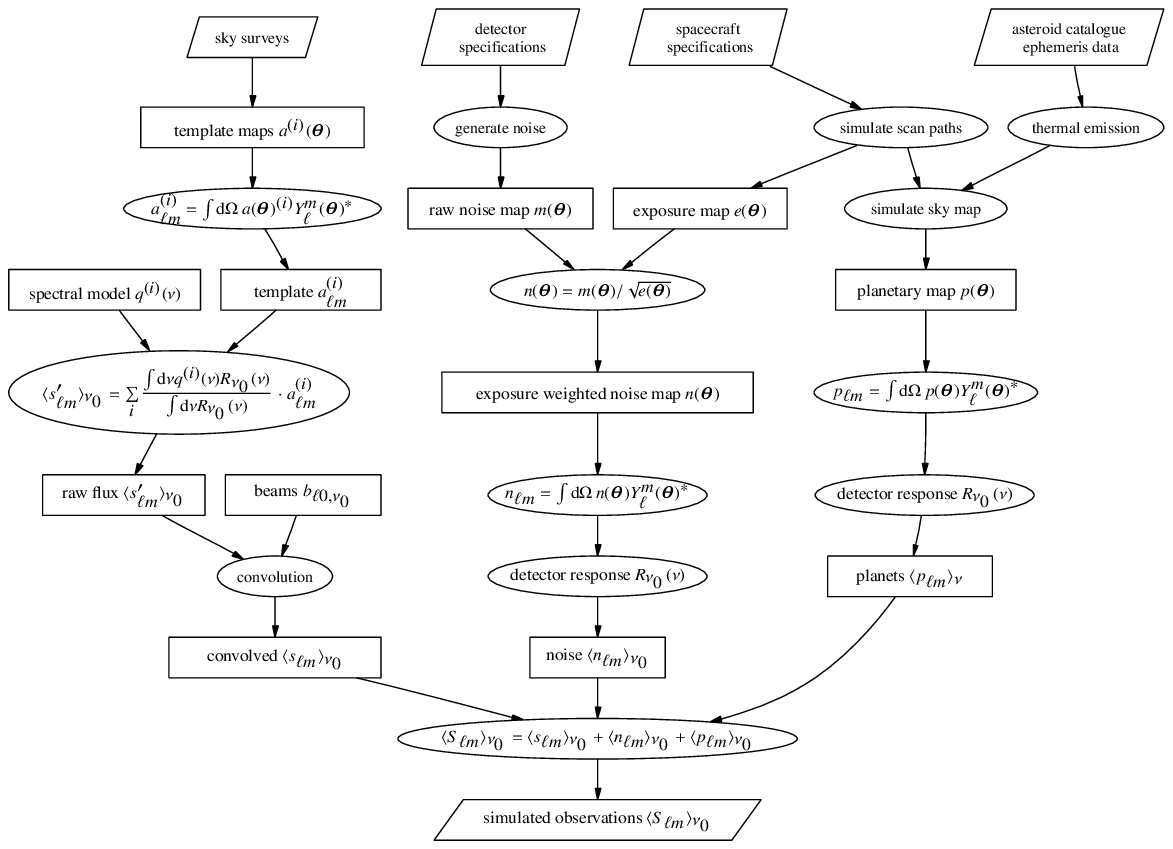}} &
\hspace{-0.5cm}
\resizebox{!}{9.5cm}{\includegraphics{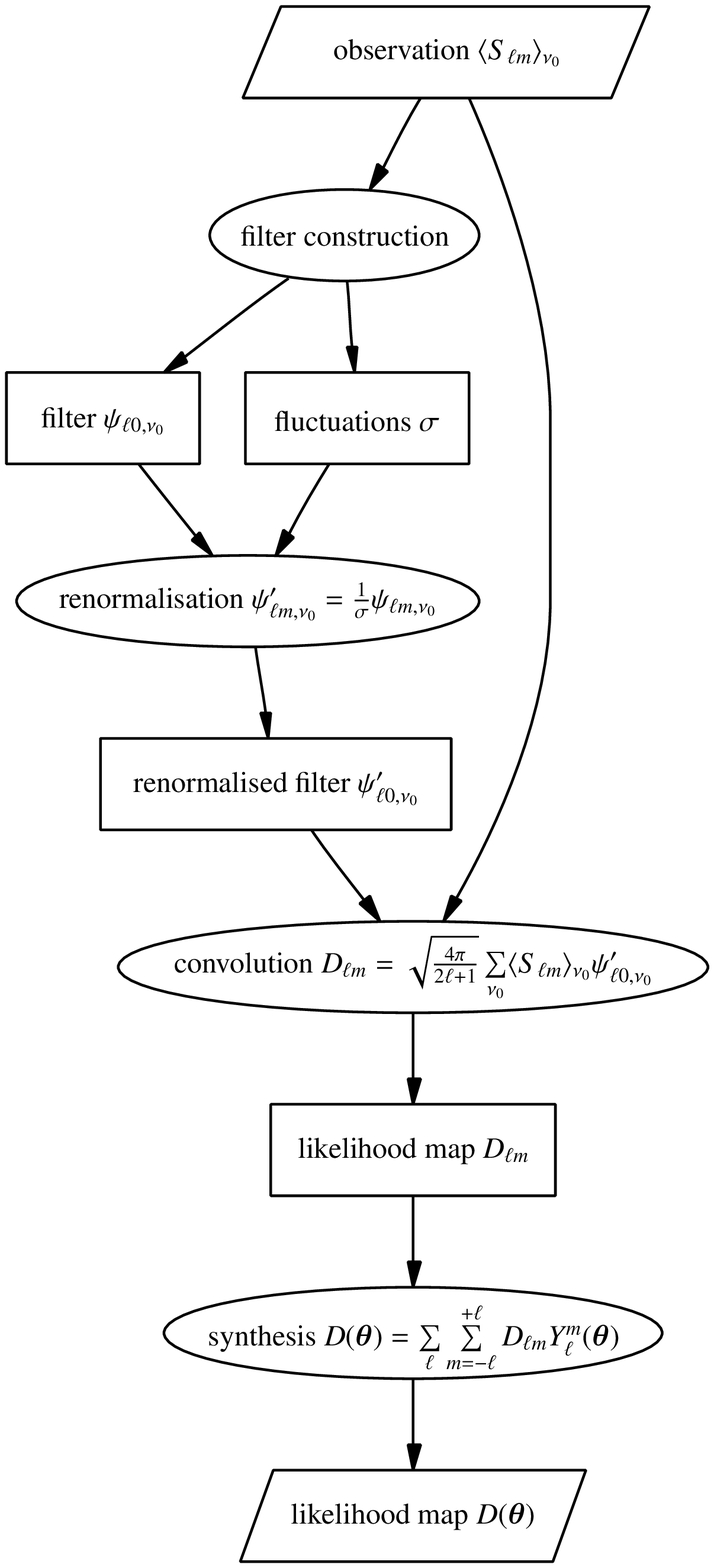}}
\end{tabular}
\caption{Flow chart summarising all steps involved in the simulation of \planck-observations and the derivation of $\bra 
S_{\ell m}\ket_\nu$-coefficients (left panel) and in the filter construction and signal extraction (right panel).}

\label{figure_flowchart_analysis}
\end{figure*}

It should be emphasised that the filters work exceptionally well despite the fact that the Milky Way clearly is a non-Gaussian 
feature, whereas Gaussianity of the fluctuating background was an important assumption in the derivation of the filter kernels. 
Furthermore, the filters sucessfully separate and amplify the weak SZ-signal in the presence of seven different noise 
contributions (CMB, four Galactic foregrounds, thermal emission from bodies of the Solar system and instrumental noise) that 
exhibit different spectral behaviours by relying on just nine broad-band measurements. Fig.~\ref{figure_flowchart_analysis} 
summarises all steps involved in the simulation of \planck-observations, filter derivation and signal extraction.

\section{Summary and conclusion}\label{sect_summary}
A simulation for assessing \plancks~SZ-capabilities in the presence of spurious signals is presented that combines maps 
of the thermal and kinetic SZ-effects with a realisation of the cosmic microwave background (CMB), in addition to 
Galactic foregrounds (synchrotron emission, free-free emission, thermal emission from dust, CO-line radiation) as well 
as the sub-millimetric emission from celestial bodies of our Solar system. Additionally, observational issues such as 
the finite angular resolution and spatially non-uniform instrumental noise of \planck~are taken into account.

\begin{itemize}
\item{Templates for modelling the free-free emission and the carbon monoxide-line emission have been added to the 
\planck-simulation pipeline. The free-free template relies on an $H_\alpha$-survey of the Milky Way. The spectral properties of 
both foregrounds are modelled with reasonable parameter choices, i.e. $T_p=10^4$~K for the free-free plasma temperature and 
$T_\mathrm{CO}=20$~K for the mean temperature of giant molecular clouds.}

\item{An extensive package for modelling the sub-millimetric emission from planet and asteroids has been implemented for 
\planck, that solves the heat balance equation of each celestial body. It takes the movement of the planets and asteroids into 
account, which causes, due to \plancks scanning strategy, double detections separated by approximate half-year intervals. The 
total number of asteroids implemented is $\simeq 1200$.}

\item{The foregrounds have been combined under proper inclusion of \plancks frequency response windows in order to yield a set 
of flux maps. The auto- and cross-correlation properties of those maps are investigated in detail. Furthermore, their 
decomposition into spherical harmonics $\bra S_{\ell m}\ket_\nu$ serve as the basis for the filter construction. It should be 
emphasised that the spectral properties of a foreground component were assumed to be isotropic.}

\item{In order to separate the SZ-Signal and to suppress the foreground components, the theory of matched and scale-adaptive 
filtering has been extended to spherical data sets. The formulae in the context of spherical coordinates and 
$Y_{\ell}^m$-decomposition are analogous to those derived for Cartesian coordinate systems and Fourier-transforms. }

\item{The global properties of filter kernel shapes are examined as functions of observing channel, composition of noise, 
parameters of the profile to be detected and spectral dependence of the signal. Transformation of the filter kernels to real 
space yields functions that resemble the Mexican-hat wavelets, but show more than one oscillation. The shape of the filter 
kernels can be understood intuitively: They subject the maps to high-pass filtering while retaining structures similar in 
angular extent to the predefined profile size. The signal is then amplified by linear combination of the maps, which again 
is apparent in the sign of the filter kernels.}

\item{The functionality of the filtering scheme is verified by applying them to simulated observations. It is found that the 
Galactic foregrounds can be suppressed very effectively so that the SZ-cluster signals can be retrieved. Comparing the two 
filters, the scale-adaptive filter performs not as good as the matched filter, which is in accordance to the findings of 
\citet{2002MNRAS.336.1057H, 2002ApJ...580..610H}. It should be emphasised that for the derivation of the filter kernels 
nothing but a model profile and all cross-power spectra (in \plancks case a total number of 45 independent 
$C_{\ell,\nu_1\nu_2}$-sets) are used.}
\end{itemize}

The scientific exploitation of our simulation and the characterisation of \plancks SZ-cluster sample, e.g. the number density 
as a function of detection sigificance as well as filter parameters, spatial distribution in dependence on Galactic and 
ecliptic latitude and the distribution in redshift, mass and apparent size will be the subject of a forthcoming paper.

\section*{Acknowledgements}
The authors would like to thank Torsten En{\ss}lin for careful reading of the manuscript. The support provided by Martin 
Reinecke in enhancing the \planck-simulation tools and adding custom changes is greatly appreciated.

\appendix

\section{Optimised filter for single frequency all-sky observations}\label{appendix_sphsingle}
This appendix derives the optimised filters for single frequency all-sky observations and thus serves as a detailed 
supplement to Sect.~\ref{sect_filtering} where optimised filters for multi-frequency observations were derived. The formalism 
outlined in this appendix might be applied to future all-sky surveys in the X-ray or microwave regime.

\subsection{Assumptions and definitions}
In order to construct our filters, we consider an all-sky map of the detected scalar field $s(\btheta)$
\begin{equation}
  \label{eq:app:signal}
  s(\btheta) = y(|\btheta-\btheta_0|) + n(\btheta), 
\end{equation}
where $\btheta = (\vartheta, \varphi)$ denotes a two-dimensional vector on the sphere and $\btheta_0$ is the source location.  
The first term of the right-hand side represents the amplitude of the sources to be detected, while the second term in 
corresponds to the generalised noise present in the map which is composed of any detected features other than the desired 
signal including for instance instrumental noise. The statistical properties of the noise are assumed to be characterised by 
its power spectrum $\left\bra n_{\ell m} n^*_{\ell' m'} \right\ket = C_{\ell} \delta_{\ell \ell'} \delta_{m,m'}$. In order to 
sketch the construction of the optimised filter, we assume an individual cluster situated at the North pole 
($\btheta_0=\bld{0}$) with a characteristic angular SZ-signal $y(\theta = |\btheta|) = A \tau(\theta)$, separating the 
amplitude $A$ and the profile $\tau(\theta)$.  



\subsection{Convolution theorem on the sphere}\label{appendix_convolution}
Filtering a scalar field on the sphere with an arbitrary, asymmetric kernel requires the specification of the convolution path 
as well as the orientation of the filter kernel at each position on the sphere in order to apply any convolution algorithm 
\citep{2001PhRvD..63l3002W}. Because of the simplifying restriction to centrally symmetric filter kernels, we give the theorem 
for the convolution of two functions, one of which is assumed to be centrally symmetric. The filtered field 
$u(\bbeta)$ is obtained by convolution of the centrally symmetric filter function $\psi(\theta)$ with the scalar field on the 
sphere $s(\btheta)$,
\begin{equation}
  \label{eq:app:convolve1}
  u(\bbeta) = \int \dd \Omega\, s(\btheta) \psi(|\btheta-\bbeta|).
\end{equation}
Expansion of these two scalar fields into spherical harmonics yields
\begin{eqnarray}
  \label{eq:app:convolve2}
  s(\btheta) &=& \sum_{\ell=0}^\infty \sum_{m=-\ell}^{+\ell} s_{\ell m}^{}\,
                 Y_\ell^m(\btheta), \\
  \psi(\theta) &=& \sum_{\ell=0}^\infty \sum_{m=-\ell}^{+\ell} \psi_{\ell m}^{}\,
                   Y_\ell^m(\theta)
                =  \sum_{\ell=0}^\infty \sqrt{\frac{2\ell+1}{4\pi}}
                   \psi_{\ell 0}^{}\,P_\ell^{}(\cos\theta).
\end{eqnarray}
The last step assumes central symmetry. In this case, only modes with $m=0$ are contributing. For proceeding,
the addition theorem for Legendre polynomials $P_\ell(x)$ \citep{1995mmp..book.....A} is used in substituting 
$\gamma=|\btheta-\bbeta|$:\begin{equation}
  \label{eq:app:at}
  P_\ell(\cos\gamma) = \frac{4\pi}{2\ell+1}\sum_{m=-\ell}^{+\ell} Y_\ell^m(\btheta)\,Y_\ell^{m*}(\bbeta).
\end{equation}
Combining these equations and applying the completeness relation yields the convolution relation for a centrally symmetric 
filter kernel,
\begin{equation}
  \label{eq:app:ct}
  u(\bbeta) = \sum_{\ell=0}^{\infty}\sum_{m=-\ell}^{+\ell} u_{\ell m}^{}  Y_\ell^m(\bbeta),
  \quad\mbox{with}\quad
  u_{\ell m} = \sqrt{\frac{4 \pi}{2 \ell+1}} s_{\ell m}\,\psi_{\ell 0} \,.
\end{equation}

\subsection{Concepts of optimised filtering on the sphere}\label{appendix_sphere_filter}
The idea of optimised matched filters was proposed by \citet{1998ApJ...500L..83T}, and generalised to scale-adaptive filters by 
\citet{2001ApJ...552..484S} for a flat topology. The construction of a centrally symmetric optimised filter function 
$\psi(\theta)$ for the amplification and detection of signal profiles differing only in size but not in shape implies a 
family of filters $\psi(\theta/R)$ introducing a scaling parameter $R$. Decomposing the family of filter functions 
$\psi(\theta/R)$ into spherical harmonics yields

\begin{eqnarray}
  \label{eq:app:filter}
  \psi\left(\frac{\theta}{R}\right) &=& R^2\sum_{\ell=0}^\infty
  \sqrt{\frac{2\ell+1}{4\pi}}\psi_{\ell 0}(R)\,P_\ell^{}(\cos\theta),\\
  \psi_{\ell 0}(R) &=& \frac{1}{R^2} \int \dd^2\theta\, 
  \sqrt{\frac{2\ell+1}{4\pi}}\psi\left(\frac{\theta}{R}\right)
  P_\ell(\cos\theta),
\end{eqnarray}
while allowing for central symmetry of the filter function.  For a particular choice of $R$ the filtered field $u(R,\bbeta)$ is 
obtained by convolution (c.f. Appendix.~\ref{appendix_convolution}):
\begin{eqnarray}
  \label{eq:app:udef}
  u(R,\bbeta) & = & \sum_{\ell=0}^{\infty}\sum_{m=-\ell}^{+\ell} u_{\ell m}^{}  Y_\ell^m(\bbeta),
  \quad\mbox{and}\\
  u_{\ell m} & = & \sqrt{\frac{4 \pi}{2 \ell+1}} s_{\ell m}\,\psi_{\ell 0}(R) \,.
\end{eqnarray}
Taking into account the vanishing expectation value of the noise $\bra n_{\nu}(\btheta) \ket = 0$, the expectation value of the 
filtered field at the North pole $\bbeta=\bld{0}$ is given by

\begin{equation}
  \label{eq:app:umean}
  \bra u(R,\bld{0})\ket = 
  A \sum_{\ell=0}^{\infty} \tau_{\ell 0}\,
  \psi_{\ell 0}(R).
\end{equation}
Assuming that the power spectrum of the signal is negligible compared to the noise power spectrum, the variance of the filtered 
field is given by
\begin{equation}
  \label{eq:app:uvariance}
  \sigma_u^2(R) = \left\bra \left[u(R,\bbeta) - \bra u(R,\bbeta)\ket\right]^2\right\ket 
  =\sum_{\ell=0}^{\infty} C_{\ell}^{}\,\psi_{\ell 0}^2(R).
\end{equation}

While the optimised {\em matched filter} in the case of single frequency observations is defined to obey the first two of
the following conditions, the optimised {\em scale-adaptive filter} is required to obey all three conditions:

\begin{enumerate}
\item{ The filtered field $u(R,\bld{0})$ is an unbiased estimator of the source amplitude $A$ at the true source position, 
i.e. $\bra u(R,\bld{0})\ket  = A$.}
\item{The variance of $u(R,\bbeta)$ has a minimum at the scale $R$ ensuring that the filtered field is an efficient 
estimator.}
\item{The expectation value of the filtered field at the source position has an extremum with respect to the the scale $R$, 
implying  
\begin{equation}
\label{eq:app:3cond}
\frac{\upartial}{\upartial R}\bra u(R,\bld{0})\ket = 0.
\end{equation}}
\end{enumerate}

\subsection{Matched filter}
In order to derive the matched filter, constraint (i) can be rewritten yielding
\begin{equation}
  \label{eq:app:constraint1}
  \sum_{\ell=0}^\infty   \tau_{\ell 0}\, \psi_{\ell 0} = 1.
\end{equation}
Performing functional variation (with respect to the filter function $\psi$) of $\sigma_u^2(R)$ while incorporating the 
constraint (\ref{eq:app:constraint1}) through a Lagrangian multiplier yields the spherical matched filter:
\begin{equation}
  \label{eq:app:matched filter}
  \psi_{\ell 0}^{} = \alpha\, \frac{\tau_{\ell 0}}{ C_\ell}, \quad\mbox{where}\quad  \alpha^{-1} = \sum_{\ell=0}^\infty
  \frac{\tau_{\ell 0}^2}{ C_\ell}.
\end{equation}

In any realistic application, the power spectrum $C_{\ell}$ can be estimated from the observed data provided the power spectrum 
of the signal is negligible. The quantities $\alpha$, $\tau_{\ell 0}$, and thus the filter kernel $\psi_{\ell 0}$ can be 
straightforwardly computed for any model source profile $\tau(\theta)$.

\subsection{Scale-adaptive filter}
\label{sec:app:SAF}
The next step consists of reformulating constraint (iii) in order to find a convenient representation for the application of 
functional variation. The expansion coefficient of the family of filter functions $\psi(\theta/R)$ of 
eqn.~(\ref{eq:app:filter}) can be rewritten yielding
\begin{equation}
  \label{eq:app:approx}
  \psi_{\ell 0}^{}(R) = 
  \frac{1}{R^2} \int  \dd^2\theta\, 
  \psi\left(\frac{\theta}{R}\right)Y_\ell^{0}(\theta) = 
  \int  \dd^2\beta\, \psi(\beta) Y_\ell^{0}(R \beta),
\end{equation}
where $\beta \equiv \theta/R$. In general, this substitution is {\em not} valid, because $\dd^2\theta = \sin\theta\,\dd 
\theta\,\dd\phi$. In the case of localised source profiles, the angle $\theta$ is small for non-vanishing values of $\psi$ 
justifying the approximation $\sin \theta\approx \theta$. The same argument also applies for the boundaries of
integration. With the aid of eqn.~(\ref{eq:app:umean}), condition (\ref{eq:app:3cond}) reads 
\begin{equation}
  \label{eq:app:derivation1}
  \frac{\upartial}{\upartial R}\bra u(R,\bld{0})\ket =
  \sum_{\ell=0}^{\infty} \tau_{\ell 0}\,
  \frac{\upartial \psi_{\ell 0}(R)}{\upartial R} = 0.
\end{equation}
Using eqn.~(\ref{eq:app:approx}), the derivative now acts on the Legendre polynomial $P_{\ell}$, 
\begin{equation}
  \label{eq:app:derivation2}
  \sum_{\ell=0}^\infty \sqrt{\frac{2 \ell+1}{4 \pi}}\,\tau_{\ell 0}
  \int \dd^2 \beta\, \psi(\beta) 
  P_\ell' (\cos R \beta)\, \beta\,\sin R \beta = 0.
\end{equation}
Using the derivative relation of the Legendre polynomials \citep{1995mmp..book.....A},
\begin{equation}
  \label{eq:app:LegendreRelation}
  P_\ell'(x) = \frac{\ell+1}{1 - x^2}\,[x\,P_\ell(x) - P_{\ell+1}(x)],
\end{equation}
one obtains
\begin{eqnarray}
  \label{eq:app:derivation3}
  \lefteqn{\sum_{\ell=0}^\infty \sqrt{\frac{2 \ell+1}{4 \pi}}\,(\ell +1)\,
  \tau_{\ell 0}
  \int \dd^2 \beta\, \psi(\beta) \frac{R \beta}{\sin R \beta}
  \,\times}\nonumber \\ 
  &&[\cos R \beta\,P_\ell(\cos R \beta) - P_{\ell+1}(\cos R \beta)] = 0.
\end{eqnarray}
In our case, the angle $\theta$ is small for non-vanishing values of $\psi$ justifying the approximations $\sin R\beta 
\approx R\beta$ and $\cos R\beta \approx 1$. Substituting back, $\dd^2\beta = \dd^2\theta/R^2$,
introducing $x\equiv\cos \theta = \cos R \beta$, and inserting the inversion of eqn.~(\ref{eq:app:approx}), namely
\begin{equation}
  \label{eq:app:inversion}
  \psi(\beta) = \sum_{\ell'=0}^\infty \psi_{\ell'0}^{}(R) 
  Y_{\ell'0}^{0}(R \beta),
\end{equation}
one arrives at
\begin{eqnarray}
  \label{eq:app:derivation4}
  \lefteqn{\sum_{\ell',\ell=0}^\infty 
    \sqrt{\frac{2 \ell+1}{4 \pi}}\,\sqrt{\frac{2 \ell'+1}{4 \pi}}\, 
    (\ell +1)\, \tau_{\ell 0}\psi_{\ell' 0}(R)\, \times}\nonumber \\ 
  && \frac{2\pi}{R^2} \int \dd x\, P_{\ell'}(x)
     [P_\ell(x) - P_{\ell+1}(x)] = 0.
\end{eqnarray}
Applying the orthogonality relation for the Legendre polynomials,
\begin{equation}
  \label{eq:orthogonal}
  \int_{-1}^{+1}\dd x\, P_\ell (x) P_{\ell'}(x) = 
  \frac{2}{2 \ell+1} \delta_{\ell\ell'},
\end{equation}
and using the small angle approximation in the second term of eqn.~(\ref{eq:app:derivation4}) with the same argument as given 
above, yields the final result
\begin{equation}
  \label{eq:app:derivation5}
  \sum_{\ell=0}^\infty\psi_{\ell 0}(R)
  [\tau_{\ell 0} + \ell (\tau_{\ell 0} - \tau_{\ell-1, 0})] = 0.
\end{equation}
Replacing the differential quotient with the corresponding derivative is a valid approximation for $\ell\gg 1$. Thus, 
eqn.~(\ref{eq:app:derivation5}) can be recast in shorthand notation yielding
\begin{equation}
  \label{eq:app:derivation6}
  \sum_{\ell=0}^\infty\psi_{\ell 0}(R)\tau_{\ell 0} 
  \left[2 + \frac{\dd \ln \tau_{\ell 0}}{\dd \ln \ell}\right] = 0.
\end{equation}
This result could have been obtained independently by attaching the tangential plane to the North pole and applying Fourier 
decomposition of the filter function $\psi$ and the source profile $\tau$. For that reason, it is not surprising that the 
functional form of this condition on the sphere agrees with that obtained by \citet{2001ApJ...552..484S} for a flat topology in 
two dimensions.

Performing functional variation (with respect to the filter function $\psi$) of $\sigma_u^2(R)$ while interlacing the 
constraints (\ref{eq:app:constraint1}) and (\ref{eq:app:derivation6}) through a pair of Lagrangian multipliers yields the
spherical scale-adaptive filter,
\begin{eqnarray}
  \label{eq:app:scale-adaptive filter}
  \psi_{\ell 0}^{} &=& 
  \frac{\tau_{\ell 0}}{ C_\ell\,\Delta}
  \left[2 b + c - (2 a + b)\frac{\dd \ln \tau_{\ell 0}}{\dd \ln \ell}
  \right],\nonumber\\
  \Delta &=& ac-b^2, \\
  a &=& \sum_{\ell=0}^\infty 
  \frac{\tau_{\ell 0}^2}{C_\ell}, \quad
  b = \sum_{\ell=0}^\infty 
  \frac{\tau_{\ell 0}}{C_\ell}\frac{\dd \tau_{\ell 0}}{\dd \ell}, \nonumber\\
  c &=& \sum_{\ell=0}^\infty  C_\ell^{-1}
  \left(\frac{\dd \tau_{\ell 0}}{\dd \ln \ell}\right)^2.
\end{eqnarray}

As before in the case of the matched filter, the power spectrum $C_{\ell}$ can be derived from observed data provided the 
power spectrum of the signal is negligible. Assuming a model source profile $\tau(\theta)$, the quantities $\tau_{\ell 0}$, 
$a$, $b$, $c$, and finally $\psi_{\ell 0}$ can be computed in a straightforward way. The derivative of $\tau_{\ell 0}$ with 
respect to the multipole order $\ell$ is a shorthand notation of the differential quotient in eqn.~(\ref{eq:app:derivation5}).

\bibliography{bibtex/aamnem,bibtex/references}
\bibliographystyle{mn2e}

\appendix

\bsp

\label{lastpage}

\end{document}